 \documentclass[%
10pt,A4paper,twocolumn,aps,pra, superscriptaddress,longbibliography, nofootinbib,
 amsmath,amssymb,
]{revtex4-2}
\usepackage[utf8]{inputenc}
\usepackage{enumerate}
\usepackage{braket}
\usepackage{amsmath}
\usepackage{amsthm}
\usepackage{amssymb}
\usepackage{amsfonts}
\usepackage{physics}
\usepackage{dsfont}                  
\usepackage{mathtools}               
\usepackage{graphicx}
\usepackage{bbold}
\theoremstyle{plain}
 
\usepackage{float}
\usepackage[colorlinks=true, linkcolor=blue, citecolor=blue, urlcolor=blue, pdfauthor={L. Bödeker, et. al.}, pdftitle={On the interpretability of neural-network decoders}]{hyperref}
\usepackage[capitalize]{cleveref}
\usepackage{lipsum} 
\usepackage{stmaryrd} 
\usepackage{booktabs}
\usepackage{fancyhdr}
\usepackage{xcite}
\usepackage{color}
\usepackage[dvipsnames]{xcolor}
\usepackage{MnSymbol}

\usepackage{bigints}
\usepackage {stmaryrd }

\usepackage{rotating}
\usepackage{subcaption}              
\usepackage{booktabs}                
\usepackage{multirow}
\usepackage[dvipsnames]{xcolor}
\usepackage{MnSymbol}
\usepackage{comment}
\usepackage{tikz}
\usetikzlibrary{quantikz}
\usetikzlibrary{shapes.geometric}
\usetikzlibrary{positioning}
\usetikzlibrary{external}
\newtheorem{definition}{Definition}[section]
\Crefname{section}{Sec.}{Secs.}
\newcommand{\encirc}[1]{\tikz[baseline,anchor=base]{%
    \node[draw, circle, inner sep=0, minimum width=1.0em]{#1};}}

\begin{document}
\def\relv#1#2{R^{(#1)}_{#2}}

\title{On the interpretability of neural network decoders}

\author{Lukas B\"{o}deker}
\email{l.boedeker@fz-juelich.de}
\affiliation{Institute for Theoretical Nanoelectronics (PGI-2), Forschungszentrum J\"{u}lich, 52428 J\"{u}lich, Germany}
\affiliation{Institute for Quantum Information, RWTH Aachen University, 52056 Aachen, Germany}
\author{Luc J. B. Kusters}

\affiliation{Institute for Theoretical Nanoelectronics (PGI-2), Forschungszentrum J\"{u}lich, 52428 J\"{u}lich, Germany}
\affiliation{Institute for Quantum Information, RWTH Aachen University, 52056 Aachen, Germany}
\author{Markus M\"{u}ller}
\affiliation{Institute for Theoretical Nanoelectronics (PGI-2), Forschungszentrum J\"{u}lich, 52428 J\"{u}lich, Germany}
\affiliation{Institute for Quantum Information, RWTH Aachen University, 52056 Aachen, Germany}
\date{\today}

\begin{abstract}
Neural-network (NN) based decoders are becoming increasingly popular in the field of quantum error correction (QEC), including for decoding of state-of-the-art quantum computation experiments. In this work, we make use of established interpretability methods from the field of machine learning, to introduce a toolbox to achieve an understanding of the underlying decoding logic of NN decoders, which have been trained but otherwise typically operate as black-box models.
To illustrate the capabilities of the employed interpretability method, based on the Shapley value approximation, we provide an examplary case study of a NN decoder that is trained for flag-qubit based fault-tolerant (FT) QEC with the Steane code.
We show how particular decoding decisions of the NN can be interpreted, and reveal how the NN learns to capture fundamental structures in the information gained from syndrome and flag qubit measurements, in order to come to a FT correction decision. Further, we show that the understanding of how the NN obtains a decoding decision can be used on the one hand to identify flawed processing of error syndrome information by the NN, resulting in decreased decoding performance, as well as for well-informed improvements of the NN architecture. The diagnostic capabilities of the interpretability method we present can help ensure successful application of machine learning for decoding of QEC protocols.
\end{abstract}

\maketitle
\section{Introduction}
\begin{figure}
    \centering
    \includegraphics[width=1\columnwidth]
{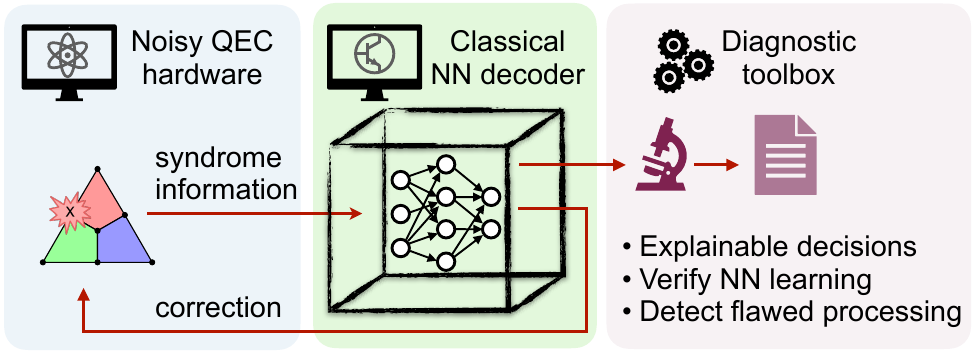}
    \caption{Opening the black box for a neural network (NN) based decoder: The interpretability toolbox offers diagnostic tools to (i) gain an understanding of how the NN decoder has come to a certain correction decision, for given error syndrome information, to (ii) supervise and verify the NN's internal learning process and to (iii)  detect flawed information processing by the decoder, which otherwise can impede fault-tolerant operation of noisy QEC codes.}
    \label{fig:black_box}
\end{figure}
Fault-tolerant (FT) quantum error correction (QEC) is a key ingredient to suppress error rates of quantum computers to a degree where scalable quantum algorithms can outperform classical computers~\cite{dalzell2023quantum}. 
State-of-the-art experiments have demonstrated QEC close to or even below break-even, below which the error rates of corrected logical qubits are reduced as compared to their physical qubit counterparts. Here, impressive breakthroughs in a variety of platforms including 
trapped ions ~\cite{Stricker2020, hilder2022fault,egan2021,Postler2022,postler2023demonstration,ryananderson2021, ryananderson2022,huang2023comparing,dasilva2024demonstration,ryan2024high,reichardt2024demonstration,Pogorelov2025}, superconducting systems~\cite{takita2017experimental,google2021exponential,Krinner2022,zhao2022realization,google2023suppressing,Gupta2024,acharya2024quantum,lacroix2024scaling,besedin2025realizing} and neutral Rydberg atom systems~\cite{Graham2022,Cong2022,Bluvstein2022,bluvstein2023logical,bluvstein2024logical,rodriguez2024experimental,bedalov2024fault} have been achieved. In these experimental realizations, the FT operation of error-corrected logical qubits is shown in different settings, demonstrating key components for large-scale error-corrected quantum computation. These active error correction protocols all have in common that partial information about erroneous processes must be processed in order to determine a recovery operation that -- if successful -- yields the preservation of the logical information.
In error correcting (topological) stabilizer codes~\cite{gottesman_stabilizer_1997}, the logical state information of a single qubit is encoded non-locally across many qubits. To detect errors, local parity check operators are measured, which leave logical information untouched. The according measurement outcomes form the error syndrome and can be used to determine a recovery operation to revoke the error with a certain probability.
The computational cost of decoding of the syndrome scales, in general, exponentially with growing code-size, when performing optimal decoding that yields the highest success probability of revoking the error (maximum likelihood decoding \cite{sundaresan2203matching}). For this reason, depending on the code, there exists a plethora of efficient decoders of suboptimal performance that aim at approximating optimal decoding while lowering the computational effort.
To illustrate the vibrant field of finding efficient decoding algorithms, we point out a number of works for two topological code families of surface~\cite{Dennis2002} and color codes~\cite{Bombin2006}. These decoding approaches~\cite{fowler2013minimum,bravyi2014efficient,duclos2010fast,higgott2023improved,Delfosse2022UFD,delfosse2023splitting,delfosse2014decoding,chamberland2020triangular,kubica2023efficient,Miguel2023,Rodriguez2022,berent2023decoding} are not meant to be a complete list.

Neural network (NN) based decoders form a versatile decoding paradigm, applicable to a wide range of quantum correcting codes and FT quantum computation protocols~\cite{Torlai2017,Varsamopoulos2017, Krastanov2017, Baireuther2018,Chamberland2018p2,varsamopoulos2019comparing, Baireuther2019, varsamopoulos2020decoding,andreasson2019quantum,Davaasuren2020,meinerz2022scalable,varbanov2023neural,bausch2023learning,cao2023qecgpt,Bordoni2023,Ninkovic2024GraphDecoding,maan2024machine}.
These decoders have the advantage of being able to adapt autonomously to the different noise models of any underlying physical implementation through learning, which can be based on numerical simulation or experimental data~\cite{Bordoni2023,varbanov2023neural,bausch2023learning}.
Furthermore, trained NN decoders have shown to be capable of generalizing to different decoding situations beyond those they have been trained for, such as,~higher code distances or different numbers of syndrome extraction repetitions~\cite{varsamopoulos2019comparing,Baireuther2019}.
On the other side, the performance of these decoders is heavily dependent on choosing an appropriate NN model and the success of its optimization (training). The latter in turn depends on the availability of training data and computational resources for training. The training of a NN becomes in general a resource-intensive endeavor with growing size of the NN model. For a readily trained NN on the other hand, determining a decoding decision is typically straight-forward and not computationally intensive. Indeed, trained NN decoders have shown good decoding performance whilst showing small overhead in terms of classical computations~\cite{varsamopoulos2020decoding, Chamberland2018p2}.

A caveat regarding the use of NN decoding is that NN decoders typically act as a black-box,~\cref{fig:black_box} meaning that the input-output relationship of the neural network is not accessible in the same way as it would be for a human-designed decoding algorithm. In other words, it is in the general case hard to retrace how the NN obtains its decoding result from the input information.
This lack of structural understanding of the decision model when employing NNs is a shortcoming, which might potentially limit the applicability of this approach, in particular when operating larger QEC codes under realistic experimental noise and when executing more complex tasks such as error correction of logical quantum algorithms. Furthermore, the black-box type decision character of the decoding process could screen non-optimal performance or the reasons for a failed training of the NN decoder.

In this work, we investigate interpretability methods for NNs to open this black-box, as schematically shown in~\cref{fig:black_box}. The techniques we employ are adapted from the field of explainable machine learning, often referred to as Explainable Artificial Intelligence (XAI)~\cite{LundbergLee2017,MolnarXAIguid,Bach2015}, which is a sub-field of machine learning~\cite{Goodfellow-et-al-2016} focusing on increasing the transparency and interpretability of machine learning models. The goal here is to provide insights into how machine learning models make predictions and decisions, in order to check for plausibility and to detect possible malfunctions of the model~\cite{Bach2015,LIME2016,Doshi2017, Montavon2018, LundbergLee2017, Montavon2018,Montavon2019, Dhamdhere2020shapley,IntegratedGradients, SmoothGrad}. In XAI a distinction is made between local explanation methods~\cite{LIME2016, LundbergLee2017, IntegratedGradients, SmoothGrad, Montavon2018, Montavon2019, Bach2015, LundbergLee2017} and global explanation methods~\cite{Doshi2017}. Local explanation methods aim to explain individual predictions, while global explanation methods aim to provide insights into the workings of the model as a whole. Global explanations are often obtained by aggregating over many local explanations~\cite{Montavon2018}.

In our work, we will focus on a model-agnostic and local method, the so-called Shapley values~\cite{Shapley1953,DeepLIFT2017} for interpreting individual decoding outcomes of a NN decoder. These interpretations are given in terms of an importance score of sub-sets of the input which in our case are parts of the syndrome information.
From the individual decoding interpretations, we can derive global conclusions about the decoder working by analyzing the statistics of the interpretations.
This Shapley-value based method that we adapt for the field of NN decoding is general in the sense that it can be applied for any kind of NN architecture that supports backpropagation and any underlying QEC decoding task.
We show that our interpretability method can be used to certify the learning success of a NN to perform FT decoding. Interestingly, this learning success can be temporally resolved along the training of the NN. This  allows one to spectate the learning transition of the NN, first learning simple non-FT decoding strategies, and later in the training process the discovery and adoption of a FT decoding behaviour.
Furthermore, we can infer whether the NN has understood how to utilize correlations between $X$ and $Z$ syndromes, relevant for the correction of phase and bit flip errors, respectively, to refine the decoding decisions it is making.
Apart from applying the interpretability method to check for desired properties, we show that the employed interpretability analysis also enables the identification of unwanted functional behaviour that the NN can acquire during the training, and which deteriorates the decoding performance and can be attributed to overfitting. Such diagnostic element then allows, in a general setting, for an informed augmentation of the NN decoder to ultimately improve the logical performance of the underlying error correction protocol.


For concreteness, we test our interpretability method based on numerical simulations of a flag-FT~\cite{Chamberland2018,Chamberland2020,Chao2018} circuit implementation of a QEC for the distance-3 Steane code~\cite{Steane96_7qbcode,Bombin2006, bombin2013} under circuit level noise in a quantum memory setting. This choice is motivated by posing the learning challenge for the NN decoder to combine information from syndrome as well as from flag qubits, obtained from various measurement rounds. If this combination of information pieces is successfully learned by the NN, it should then be able to correct for any single fault that occurs during the stabilizer readout, and thereby maintain the FT of the QEC code in combination with the suitably constructed flag-qubit based syndrome readout quantum circuitry. Furthermore, as for other CSS codes, in this setting it is possible to study the effect of the correlation between $X$ and $Z$ errors for the decoding decision as the respective two decoding problems can be solved independently. Additionally, studying this concrete setting is also motivated by its near-term experimental relevance~\cite{Bluvstein2022,bluvstein2023logical,Postler2022,postler2023demonstration,ryananderson2022}.
In terms of the NN model we choose for decoding, we work with simple recurrent NNs, as previously used in the QEC decoding literature~\cite{Baireuther2019}. These NNs are employed and trained to classify the logical-flip parity on the QEC code after $T$ rounds of faulty syndrome measurement.




This manuscript is structured as follows: In Sec.~\ref{sec:Neural Network Decoding} we explain the NN decoding paradigm and introduce the exemplary QEC setting, which we later test on our interpretation method. In Sec.~\ref{sec:NN decoder interpretation}, we introduce the Shapley value, outline how it can be calculated and explain how it can be used to derive local as well as global interpretations. In Sec.~\ref{sec:Monitoring the learning of Fault Tolerance}, we then present how the learning progress of the NN decoder is analyzed, focusing on the NN's ability to decode fault-tolerantly. In Sec.~\ref{sec:Diagnosing NN malfunction for the two headed NN} we show, based on a fully trained model, how possible malfunctions of the NN decoder can be uncovered. Finally, Sec.~\ref{sec:Conclusion and Outlook} provides conclusions and an outlook to potential extensions of the work.

\section{Neural Network Decoding}
\label{sec:Neural Network Decoding}

\begin{figure*}
    \centering
    \includegraphics[width=1.75\columnwidth]{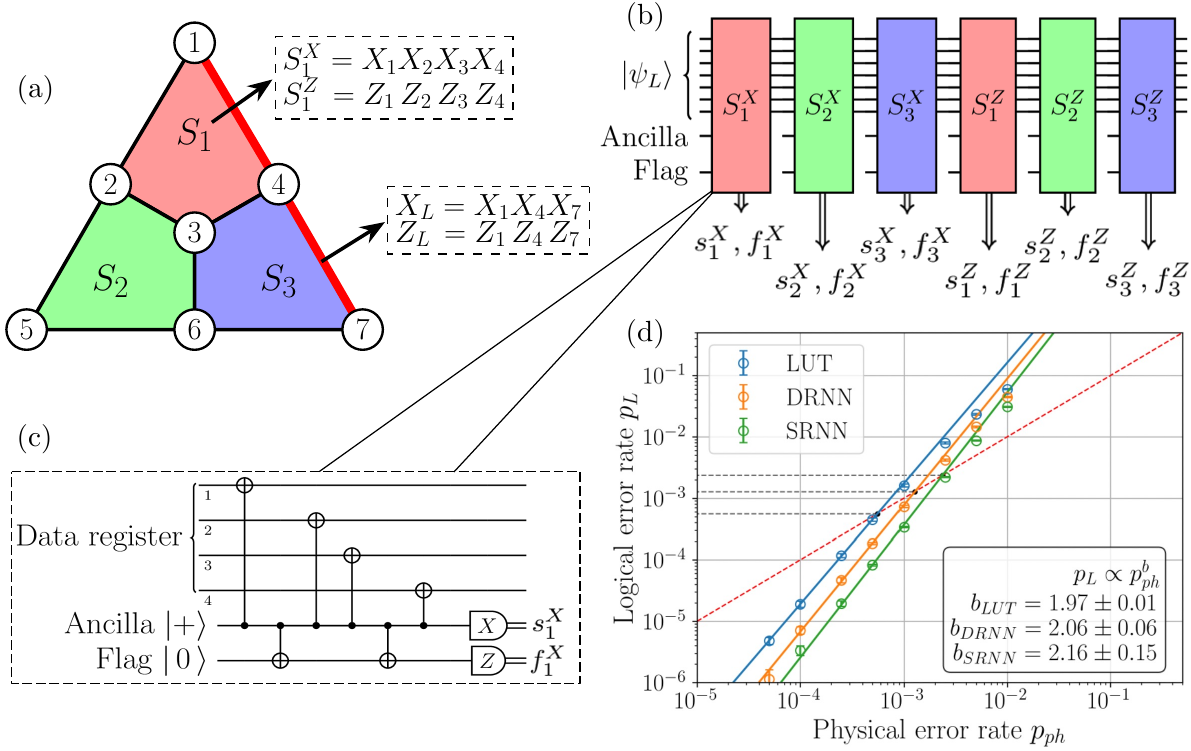}
    \caption{
    (a) The $\llbracket 7,1,3\rrbracket$ color code. The vertices represent data qubits and the faces represent the $X-$ and $Z-$type stabilizer generators. The pair of red plaquette stabilizer generators is indicated as an example, as well as a representation of the logical operators. The other four stabilizer generators are defined analogously as the green and blue plaquette. (b) Measurement schedule for the flagged stabilizer readout. The measurements return the stabilizer parity and flag information. (c) Exemplary flag-based fault-tolerant measurement circuit for the $S^X_1$ stabilizer. The flag qubit that is coupled to the ancilla qubit is capable to identify any potentially dangerous error propagations. (d) Performance comparison of three fault-tolerant decoding schemes: look-up table decoder (LUT in blue), dual output recurrent neural network decoder (DRNN in orange) and single output recurrent neural network decoder (SRNN in green).  All decoders show quadratic scaling of the logical error rate $p_L$ with the physical error rate $p_{ph}$, implying the capability of the decoders to operate fault-tolerantly and correct all single faults in the circuit. Error bars are computed as the one-sigma Wilson interval, see~\cref{subsec:wilson}.}
    \label{fig:steane}
\end{figure*}
\begin{figure}
    \centering
    \includegraphics[width=\columnwidth]{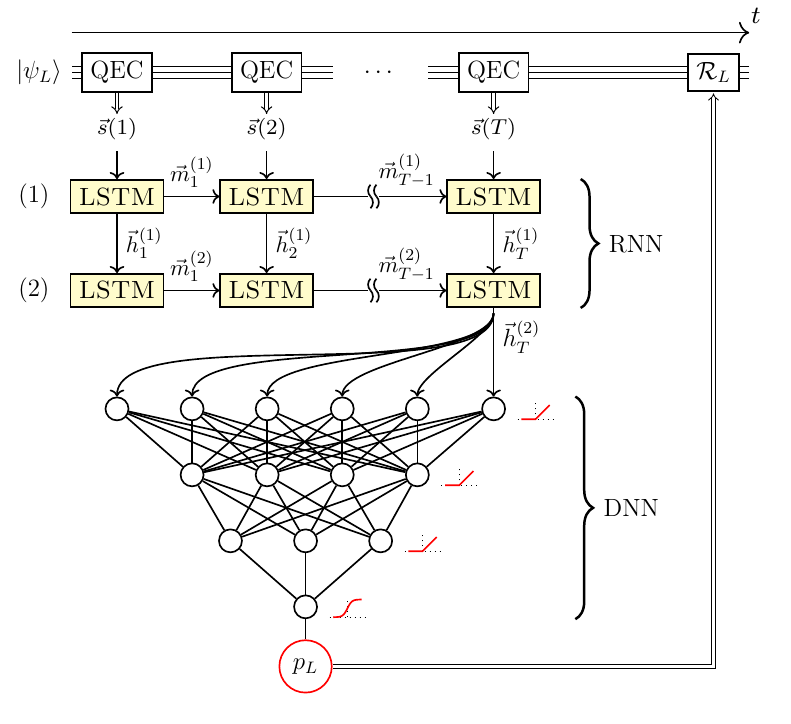}
    \caption{Recurrent neural network (RNN) architecture for logical error correction given a syndrome-flag volume based on ~\cite{Baireuther2019}. All stabilizer parities and flags are measured sequentially (QEC block), and the network passes the respective measurement outcomes through two RNN layers using the LSTM architecture. The interlayer messages $\vec m_i^{(l)}$ consist of an internal `cell state' $\vec c_i^{(l)}$ combined with the messages $\vec h_i^{(l)}$. This is followed by a dense neural network (DNN) for post-processing. The latter network consists of multiple layers of neurons, each equipped with an activation function, indicated schematically by the small red graphs. All but the output neurons have a rectified linear unit (ReLU) as activation function, and the output neuron is equipped with a smooth sigmoid function. The output is the predicted logical flip probability. If a flip is detected, a recovery operation $\mathcal{R}_L$ is performed by applying the corresponding logical operator.}
    \label{fig:RNN}
\end{figure}

Various NN architectures and approaches for decoding have been explored~\cite{Torlai2017,Chamberland2018p2}. Given a QEC experiment with $T$ rounds of repeated stabilizer measurements, one can for instance provide the whole space-time syndrome volume~\cite{Fowler2012} as input to the NN to perform the decoding. For growing code distances and therefore large spatial syndrome, these NNs can further be enriched by sparsely connected network parts as elements of their architecture to perform a spatially local preprocessing, as it is done in convolutional NNs~\cite{Davaasuren2020,meinerz2022scalable,Bordoni2023}.
Still, these networks have the caveat that, after training, they are fixed to a single QEC protocol together with a syndrome volume of fixed size as input. If one is, however, faced with a protocol where a variable number of stabilizer measurement rounds is needed, and for which the circuitry that is run is time-translationally invariant, it is more convenient being able to pass the syndrome information to the NN sequentially, in a time-resolved way~\cite{Baireuther2019,varbanov2023neural,bausch2023learning}. Such a procedure is reminiscent of algorithmic decoders that employ a sliding window to process the syndrome history~\cite{huang2023improved,skoric2023parallel}.  In order to be able to still use information of stabilizer measurements at different times, the NN is required to be equipped with a memory~\cite{Hochreiter2001} from which an overall decoding decision can be determined considering the joint syndrome volume information.

The problem of decoding can be formulated in different ways for a NN that is to be trained. A microscopic approach would be to train a NN to determine every fault that has occurred given the syndrome history. A simplified, but for decoding sufficient, task for the NN could be to only predict the logical parity of the accumulated errors. If a NN can predict this logical parity reliably, one could assume that it processes the syndrome information correctly into information about the errors internally.
We employ a type of recurrent neural network (RNN) that can process a variable amount of the syndrome readouts, similar as in the works~\cite{Varsamopoulos2017, Baireuther2019, Chamberland2018p2}. The RNN is trained to predict the logical parities after a variable number of $\{T_i\}$ rounds of stabilizer measurement such that after the training it will be able to make this prediction after being fed measurement outcomes from $T'\nin \{T_i\}$ rounds of stabilizer measurement~\cref{fig:RNN}, see~\cref{sec:RNN_architecture} for details about the network specifications.

\emph{Quantum memory experiment---}
This setting is relevant if one wants to protect the information encoded in a logical qubit state $\ket{\psi_L}$ that is left to idle, while a number $T$ of stabilizer measurement cycles is performed, of which one is shown in~\cref{fig:steane} $(b)$ for the Steane code.

For the Steane code, the problem of decoding amounts to a classification problem, where as input the joint vector of syndrome and flag measurements $\vec s$ encodes partial information about where in the circuit faults have occurred. The task of the NN is therefore to decide on a logical correction after $T$ rounds of stabilizer measurements, see~\cref{fig:RNN}. To evaluate the success of the logical correction that is proposed by the NN, the logical state needs to be brought back to the code space. The latter can be done by a simple correction based on the data qubit readout after the $T$ rounds of stabilizer measurement.

In more formal terms, the decoding is broken up according to the decomposed error $E$ as $E = S \cdot P \cdot P_L$, where $S$ is an element of the stabilizer group, $P_L$ is a logical operator (including the identity operator $I^{\otimes n}$), and $P$ is a `pure error' taking the state out of the logical subspace. The latter is chosen to have the smallest possible support, implicating that no logical operator can be contained in $P$. As an example based on the Steane code layout in~\cref{fig:steane} $(a)$, consider the error $E=X_1X_4X_5X_7$ which can be written as $E=I\cdot X_5\cdot X_L$, where one representation of the logical error is $X_L=X_1X_4X_7$ and the pure error that causes a nontrivial syndrome is given as $P=X_5$.
Given only the syndrome information of this pure error $P=X_5$, a decoder would assert a correction $C=X_5$. Overall, this means that after correction, we are still left with an uncorrected logical error $E\cdot C=X_L$ and therefore a failure of the memory.
Let us assume now that the decoder has access to the syndrome history of $T$ rounds of stabilizer measurements and that the error $E=X_1X_4X_5X_7$, is the cumulative error that the data qubits acquired during this time.
The correction of the error may now be split up as $C=C_L\cdot C_P$, where the 'logical correction' $C_L=X_L^{n=0,1}Z_L^{m=0,1}$ can be determined after correcting the pure error with $C_P$. The correction of the pure error is conducted based only on the stabilizer violations after the $T$ rounds of syndrome extraction, such that the logical state is brought back to the code space, i.e. $[C_P\cdot P,S]=0$ for all elements of the stabilizer group $S$. In the example case, we would still get $C_P=X_5$ as the simple correction, bringing the state back to the code space, $P\cdot C_P=I^{\otimes n}$.
In general, the correction of the pure error can however also yield a logical operator. As a consequence, the desired, successful logical correction, can be written as $C_L=S'\cdot P_L \cdot (C_P P)$.
For the remaining error $E\cdot C_P=X_1X_4X_7$, the desired logical correction by the decoder based on the syndrome history would be $C_L=X_L$.

Again, by this treatment, determining $C_L$ turns into a classification of whether a logical bit- or phase-flip has occurred or not, based on the full syndrome history. Consequently, the logical parities can be represented as two binary values, which can be predicted using a neural network that is fed with the syndrome flag history. In our simulations, the correction of this pure error $C_P$ is performed by a virtual error correction step based on a perfect final data qubit readout, from which a syndrome can be derived that can be corrected by means of look-up-table.

\emph{Fault-tolerant stabilizer measurement---}
We consider depolarizing circuit level noise~\cite{tomita2014low} with a physical error parameter $p_{ph}$. Specifically, we assume noisy qubit initialization and measurement where the outcome is inverted with a probability of $2/3\,p_{ph}$. After every single-qubit gate, one of the three single-qubit Pauli operators $\{X,Y,Z\}$ is applied with a probability $p_{ph}/3$. After every two-qubit gate, one of the 15 two-qubit Pauli operators $\{I, X, Y, Z\}^{\otimes2}\backslash \, I\otimes I$ is applied with a probability $p_{{ph}}/15$.
Given this noise model, a single fault on a measurement ancilla qubit can propagate onto multiple data qubits. In particular, in the Steane code no weight-two data qubit error of two Pauli operators of the same type can be corrected. If this spreading of errors was not detectable as in the case of using a single physical ancilla for the readout, the logical information would become irrecoverable. 
One example for a fault that causes a weight-two data qubit error is shown in~\cref{fig:corr_flag} $(c)$. We will call this fault class hook errors in the following. To prevent such hook errors from going unnoticed, there exist several measurement schemes which allow for syndrome extraction while preserving fault tolerance~\cite{steane1997active,shor1996proceedings,knill2005quantum,reichardt2020fault}.
The scheme of our choice~\cite{Chamberland2018} exploits an additional measurement ancilla qubit, the so-called flag, that is entangled with a syndrome ancilla during the stabilizer measurement, see~\cref{fig:steane} (c). In this way, all dangerous hook errors also propagate onto the flag qubit and cause a nontrivial measurement outcome. When such a flag is raised during one round of stabilizer measurements, all possible weight one faults can be corrected if the stabilizer measurements are repeated.

\emph{Neural network architecture---}
The recurrent cell we choose for our RNN is the so-called Long Short Term Memory (LSTM) unit~\cite{Hochreiter2001}, see~\cref{sec:RNN} for details. It is a reasonably simple and extensively tested NN element that allows for actively memorizing and forgetting information based on the input. The latter is a needed capability to combine measurement outcome information from various consecutive rounds of syndrome extraction for the decoding decision. In the overall NN, LSTM layers are complemented with non-recurrent layers. 
The embedding of the RNN as the decoder as well as its inner composition is shown schematically in Figure~\ref{fig:RNN}. The RNN receives syndrome increments and flag bits as input in a time-resolved and sequential manner. These inputs are then passed to two consecutive layers of LSTM units that allow for combining information of spatially and temporally separated syndrome bits followed by a dense feed-forward neural network (DNN) to perform post-processing. The first LSTM layer uses a many-to-many structure in space and time, i.e., it passes a sequence of outputs $\vec h_t^{(1)}\,\forall\,t$ to the next layer given a sequence of inputs $\vec{s}(t)$. On the other hand, the second LSTM layer contracts these sequences to only the final sequence output, $\vec h_{(t=T)}^{(2)}$ which is then passed to a DNN layer.
A DNN can be understood as a conventional network, consisting of layers of real-number valued neurons, layer-wise interconnected by trainable-weight matrices. By means of these trainable parameters, the neuron value of the previous layer and a non-linear function, the neuron values of the next layer are calculated.
The underlying idea of this structure is that the second-stage DNN can perform a post-processing step to compute the logical flip parity after the RNN has extracted the necessary information from the syndrome-flag time series.
The output of the joint network is the probability for the occurrence of a logical flip.
A restriction of this architecture is that it can only detect logical errors in either the $X$ or the $Z$ basis at a time, meaning that two networks have to be trained for complete decoding. We will denote this type of decoding as single-headed recurrent neural network decoding (SRNN); it will be contrasted with a network that predicts both the logical phase- and bit- flip parity in a single run. This double-headed recurrent neural network (DRNN)~\cite{Baireuther2018,Baireuther2019} and its performance will be discussed in Sec.~\ref{sec:Diagnosing NN malfunction for the two headed NN}.

\emph{Simulation of the quantum memory experiment and training ---}%
To train the network in a supervised manner, a sufficiently large set of tagged data examples i.e., tuples of (syndrome-flag-volume, logical parity) information is required. We generate such training data by the following procedure:
\begin{enumerate}
  \item A known logical state $\ket{\psi_L}$ is prepared by means of a noisy projective measurement of stabilizers. For the SRNN training, we choose $\ket{\psi_L}\in\{X_L^{m_x}\ket{0_L}\}_{m_x=0,1}$ for the bit flip decoder and $\ket{\psi_L}\in\{Z_L^{m_z}\ket{+_L}\}_{m_z=0,1}$ for the phase flip decoder. For the DRNN, we choose $\ket{\psi_L}\in\{X_L^{m_x}\ket{0_L},Z_L^{m_z}\ket{+_L}\}_{m_x=0,1;\,m_z=0,1}$.
  Each state obtains an equal amount of samples, and the logical parity information of the input state $m^{\mathrm{in}}_{x/z}\in\{0, 1\}$ is stored.
  \item A number $T$ of stabilizer measurement cycles is performed, producing a sequence of $T$ syndrome plus flag vectors, $\vec{s}(1),...,\vec{s}(T)$,~i.e., the schedule in Fig. \ref{fig:steane} (b) is repeated $T$ times.
    \item The data qubits are measured and a classical round of error correction is conducted, upon which the final logical parity $m_{\mathrm{out}}$ is well-defined.
  \item An output label for the NN training is defined as the logical flip parity $m^L_{x/z}=m^{\mathrm{in}}_{x/z} + m^{\mathrm{out}}_{x/z} \mod 2$.
\end{enumerate}
The vector of all syndrome and flag measurements, denoted as $\vec s$, is then used as the training input, whereas the logical flip parity $m^L$ may be used as the output label. After the respective NN model has converged under the training, it is tested with unseen data that can be sampled using the same procedure but was not used for the training before.

\emph{Decoding results---}
The error correction capabilities of the fully trained RNN decoders, including whether they are fault-tolerant, are presented in Fig~\ref{fig:steane} $(d)$. Here, the logical error rate per readout round of the described RNN decoders is compared with the one of a flag based sequential look-up table decoder (LUT) over a range of physical error rates.
The LUT works by sequentially reading the syndrome and flag values in the readout sequence and performing the most probable correction of fault weight one based on at most two readout round intervals. This implies that all single faults are corrected by the LUT, including hook errors, for more details see~\cref{sec:Sequential-Look-Up-Table-Decoder}. The LUT can be considered as the simplest flag-FT decoder and is used as a benchmark.
Both the LUT and the RNN decoders show a scaling of the logical error rate as $p_L\propto p_{ph}^2$ (see Fig~\ref{fig:steane} $(d)$). This indicates a fault-tolerant decoding as only two independent faults, each occurring with a probability of $\mathcal{O}(p)$ cause a logical error. The RNN decoders exhibit larger pseudo-thresholds and overall a smaller logical error rate compared to the LUT, indicating that they are capable to correct more faults of weight two or larger by exploiting information and correlations contained in the global set of measurement outcomes from the multi-round syndrome-flag stabilizer readout sequence. Hence, one can conclude that the employed NN structure is capable to process the flag information correctly and that it yields, for the present case, clearly better performance than a standard, LUT based decoding approach.

\section{NN decoder interpretability}
\label{sec:NN decoder interpretation}
\begin{figure*}
    \centering
    \includegraphics[width=1.5\columnwidth]{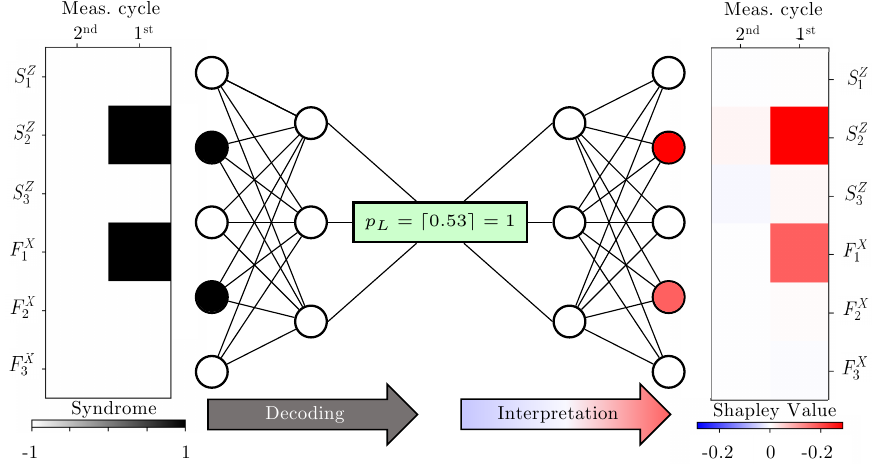}
    \caption{Minimal example of the interpretation procedure of the decoding of a two-round stabilizer measurement sequence. On the left side, the measured syndrome is fed into the NN decoder, which predicts that a hook error and therefore a logical error modulo final correction has occurred. Based on this run of the decoder (input, neuron activations), Shapley values for the input bits are calculated via backpropagation through the NN. These Shapley values are shown in color-coded form on the right side.}
    \label{fig:shap_nn}
\end{figure*}
\begin{figure*}
    \centering
    \includegraphics[width=1.6\columnwidth]{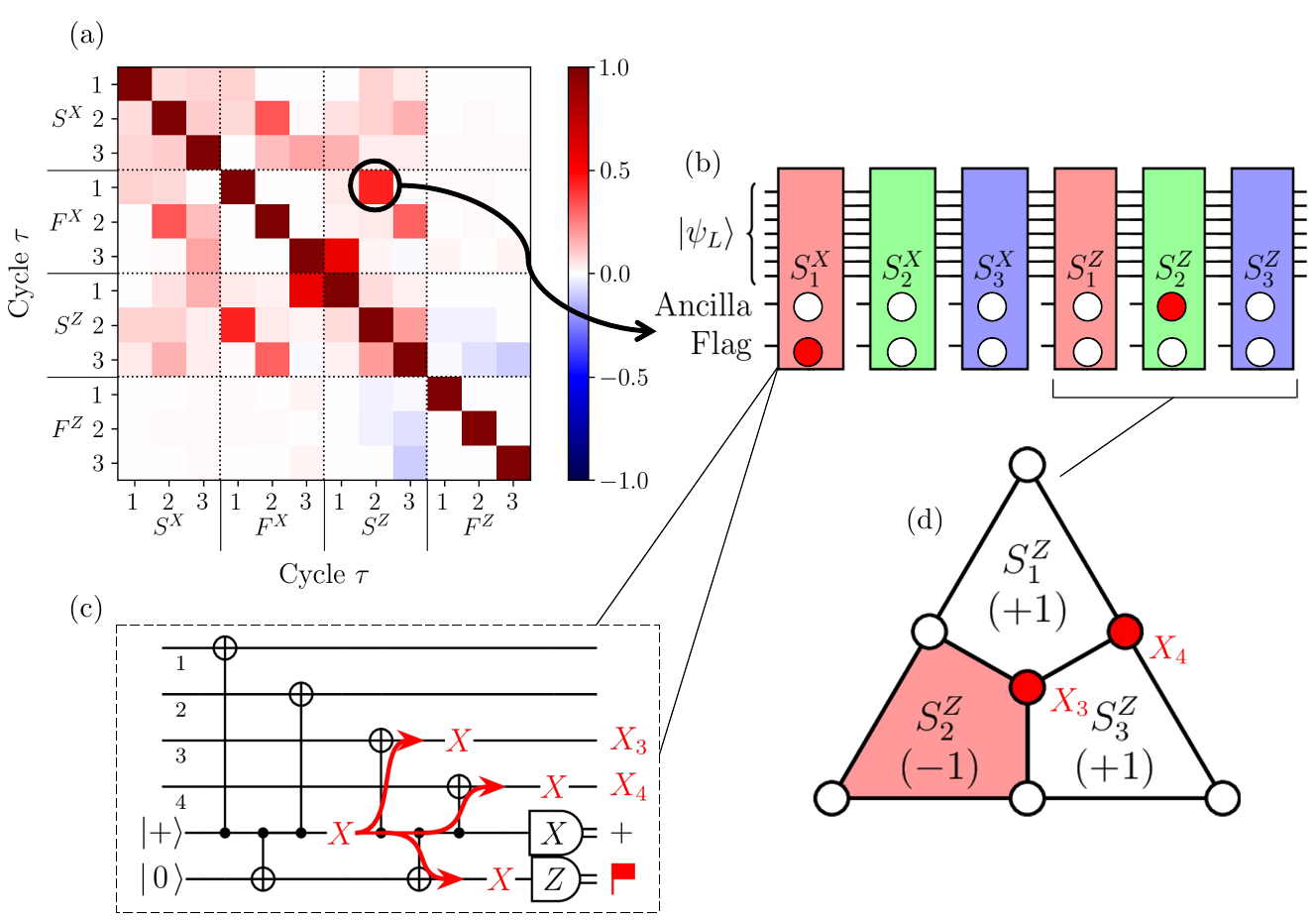}
    \caption{(a) The mutual correlation matrix of the Shapley values of syndrome and flag bits of the same QEC round ($\Delta t=0$). Each three-bit segment bundles one type of bits: syndrome parity or flag parity and $X$ or $Z$ type. The excess of correlation of the Shapley values of the stabilizer bit $S^Z_2$ and flag bit $F^X_1$ is highlighted. (b) Illustration of the corresponding hook error signature in a QEC cycle. During the first ($X$ type) stabilizer readout, a fault on the ancilla qubit takes place. The corresponding error propagation on the circuit level is as depicted in (c). Besides the error propagation onto the data qubits, it is also shown how the flag qubit picks up the error. Further, the weight 2 data-qubit error will be picked up as a $Z$-syndrome in the subsequent ($Z$ type) stabilizer readout. In (d), this $Z$-syndrome on the Steane code is shown.}
    \label{fig:corr_flag}
\end{figure*}

The problem of explaining complex models can be viewed as the challenge to create a simpler proxy-model that is accurate to a satisfactory degree. Hence, any explanation for a model can itself be seen as a model. Complex models, such as NNs, are often referred to as `black-box' models, in which the inner workings are not known or understood. Even though neural networks are built out of modular, simple, and explainable elements, their 'black-box' nature emerges due to the high degree of connectivity, non-linearity and from many degrees of freedom~\cite{HORNIK1989359,Singh2023}. For these reasons, approximations of readily trained NNs may be developed to serve as interpretability models. These explanatory models typically only capture a particular working regime of the model and are therefore only 'locally accurate', such as being an approximation around particular input values. Despite this limitation, they can still be useful to understand the underlying complex model such as a NN.

A common method of interpreting NNs is to analyze how the network transforms the input space into the output space by assigning credit to input features. Various heuristic methods have been devised to derive such credit assignment scores~\cite{LundbergLee2017,Montavon2019,zhang2021survey}. In our work, we will focus on one of these scores, the so-called Shapley value; an alternative choice is outlined in~\cref{sec:LRP}.\\

\emph{The Shapley value---}%
\label{ssub:defining-the-shapley-value}
Consider a game where a coalition of $n$ players work together to produce a shared result. The result may be assigned a numerical value, for instance, prize money in a competition. The goal is to fairly assign credit to each individual player for their contribution to achieving the shared result. To assign credit to a single player in the coalition, one analyzes how the coalition performs with that player compared to how the team would, hypothetically, perform without that player. It may be the case that some players only perform well when other players are present as well, or that certain combinations of players actually reduce the effectiveness of a team. For this reason, it is necessary to consider all possible subsets of players to accurately capture interactions between players in defining a good measure of contribution for the individual players. A solution to this problem was found by Shapley~\cite{Shapley1953}. We formally introduce the concept of the Shapley value following Ref.~\cite{Rozemberczki2022}.

To begin, we define a few quantities for discussion about $n$-player games.
\begin{definition}[Player set] Let $\mathcal{N}=\{1, 2,\ldots, N\}$ be the set of all players.  Each non-empty subset $\mathcal{S}\subseteq \mathcal{N}$ is called a coalition. The set $\mathcal{N}$ is named the `grand coalition'.
\end{definition}

\begin{definition}[Cooperative game]
  A cooperative game is defined by the pair $(\mathcal{N}, v)$ where $v: 2^{|\mathcal{N}|} \to \mathbb{R}$ is called the characteristic function, which maps any coalition $\mathcal S$ to a real number.
\end{definition}

A game in this sense is defined only by its participants and its characteristic function, without any consideration of the internal workings of the game. The characteristic function may be interpreted as the payoff of the game.
The Shapley value is a specific solution concept for the characteristic function and defined as follows:

\begin{definition}[Shapley value]\label{def:shapley-value}
  The Shapley value $\phi_i$ is a single-valued solution concept, given by
  \begin{equation}
    \phi_i(\mathcal{N}, v) = \frac{1}{|\mathcal{N}|} \sum_{\mathcal{S}\subseteq \mathcal{N}\setminus\{i\}} \begin{pmatrix}
      |\mathcal N|-1 \\ |\mathcal{S}|
    \end{pmatrix}^{-1}\underbrace{(v(\mathcal{S}\cup\{i\})-v(\mathcal{S}))}_{\text{marginal contribution}}
  \end{equation}
\end{definition}
It may be interpreted as the average marginal contribution of a player $i\in\mathcal{N}$ over all possible subsets of players $\mathcal{S}\in\mathcal{N}\setminus\{i\}$. The marginal contribution is the value a player $i$ would add to a specific coalition $\mathcal{S}$.
The Shapley value has many interesting properties, as outlined in~\cite{Rozemberczki2022} and~\cref{sec:shapley_definition_properties}, however its exact calculation is expensive as its computation cost scales as $O(2^{|\mathcal N}|)$.
When evaluating large neural networks that process many input features, it is therefore essential to come up with efficient approximations to the Shapley value. We provide references for a number of state-of-the-art approximations for universal explainability, including linear regression approaches~\cite{chen2022explaining,LundbergLee2017, frye2021shapley, covert2021improving} and Monte Carlo sampling~\cite{yuan2021explainability}, all with $O(|\mathcal N|)$ time complexity.
In this work, however, we use a neural network specific method called DeepSHAP introduced recently by Chen \textit{et al.}~\cite{chen2022explaining}, which is included as part of the open source SHAP library~\cite{SHAPGit}. It makes use of the linearization of a NN that takes place in the back-propagation procedure. Employing this linear approximation and performing a modified back-propagation, it can be shown that for expanding around an ensemble of average neuron values, the approximate Shapley value of the input features can be obtained.
For a rigorous derivation, we refer the reader to~\cite{chen2022explaining} and~\cref{sec:shapley_definition_properties}.
Note that furthermore, the approximate Shapley value of neurons in intermediate network layers can also be computed. Studying these values, however, is not part of this work, but we note that it can be an additional tool to understand or optimize a NN.

\emph{Decoder interpretation---}
Given a trained NN-based decoder and the DeepSHAP method to approximate the Shapley value efficiently, we now obtain an importance score for each syndrome and flag bit in every individual measurement sequence.
A minimal example for such an assignment of importance is illustrated in~\cref{fig:shap_nn}, where a RNN is asked to decode two rounds of flagged Steane-code stabilizer measurements. On this individual level, especially for a larger number of measurement rounds, it is possible to understand from the Shapley value distribution how the RNN regards certain syndrome(-flag) combinations to determine the logical flip parity. For instance, low Shapley values for a temporally separated pair of syndrome excitations can be read as the RNN recognizing this as a measurement error. Further, given the prediction of a logical parity flip, one can understand that  the syndrome excitations of the largest Shapley values are attributed to the faults that lead to this predicted flip.

Apart from interpreting the decoding decision based on this importance score for individual inputs, one can further interpret the global working of the NN by analyzing the statistics of Shapley values. One interesting aspect of the inner decision logic of a NN decoder is present in the context of flag-fault-tolerant stabilizer measurements during a quantum memory experiment. Here, certain combinations of nontrivial syndrome flag bit pairs are supposed to hold greater importance as they are expected to indicate a logical flip.
This suggests analyzing the statistics of Shapley values by inferring the correlation of these pairs of Shapley values corresponding to the mentioned syndrome flag bits. We calculate 
the mutual correlation matrix of Shapley values of syndrome and flag bits that are measured in the same round of syndrome extraction, i.e.~$\Delta\tau=0$ in~\cref{fig:corr_flag} (a).

Indeed, we can identify an excess in the correlation of Shapley values for syndrome-flag bit-pairs that are indicative for hook errors. Generally, there are three syndrome-flag pairings that indicate hook errors. Their Shapley-value correlation can be seen in Fig.~\ref{fig:corr_flag} (a). This is a positive verification of the NN's capability to understand the syndrome flag logic to render the propagation of dangerous errors identifiable. Note that these signatures of hook errors are observed for the bit-flip decoder in the $f_X s_Z$ Shapley correlations of the same measurement round $\Delta\tau=0$. Similarly, for the phase-flip decoder hook error signatures in the $f_Z s_X$ Shapley correlations are recognized for $\Delta \tau=1$. Generally, the specific pairings of syndrome flag bits, which are signatures for hook errors, depend on the specifics of the quantum circuitry of the syndrome flag measurement. An example of a fault propagation, and how it appears as a syndrome and flag combination during the stabilizer readout that we simulate, is illustrated in Fig.~\ref{fig:corr_flag} (b-d).

Overall, flag bits that are measured whilst the readout of stabilizers of a specific basis show the tendency to mostly correlate with syndrome bits of the respective other basis, as expected. That is to say, $f_X$ flag bits are relevant to determine logical errors mainly in combination with $s_Z$ bits and vice versa. This behavior is consistent with the possible error propagations during the stabilizer measurement, where the propagation of a hook error is shown in ~\cref{fig:corr_flag} (c). Further, the Shapley values corresponding to syndrome values of the same basis are correlated, as these natively indicate the errors on data qubits of the code. During the repeated stabilizer measurements, single data qubit errors can lead to a logical fault upon being accumulated. Therefore, to perform a good decoding, it is relevant for the NN decoder to consider pairs of syndrome values corresponding to the same stabilizer type, $Z$ or $X$, respectively.

The correlation of the importance between $X$ and $Z$ syndrome bits is less strongly pronounced. Still, it is present at a statistically significant level. This type of Shapley correlation is an indicator for the network recognizing correlations $X$ and $Z$ errors in the noise processes, which is present in our depolarizing noise model due to the presence of independent $Y$ errors. The correlation of $X$ and $Z$ errors naturally carries over to the syndrome data, where now for instance the $X$ syndrome carries partial information on the presence of $X$ errors. If the NN decoder is given the full syndrome information, one might ask whether the NN has detected such correlation at all during the training or whether it is able to make use of it. By analyzing the Shapley correlation of $X$ and $Z$ syndrome bits, this question can be answered quantitatively, as shown in~\cref{fig:corr_flag}. Hierarchically speaking, the Shapley correlations of the $X$ and $Z$ syndrome are weaker than the mutual correlation among one syndrome species. This is expected as the correlation due to $Y$ errors can only be an additional but not the main factor for determining a decoding decision. The bare presence of Shapley correlations of $X$ and $Z$ syndrome bits alone shows, however, that the NN decoder, after training, is aware of the $Y$ errors. Summarizing the hierarchy of Shapley correlations between syndrome-flag bits, the pairing that indicate hook errors are pronounced most strongly, as here single errors, if decoded incorrectly, can cause a logical flip. Other syndrome-flag pairs of the type $(S^Z,F^X)$ as well as pairing between syndromes still show modest Shapley-value correlations.

\section{Monitoring the learning of Fault Tolerance}
\label{sec:Monitoring the learning of Fault Tolerance}
\begin{figure*}
    \centering
    \includegraphics[width=1.9\columnwidth]{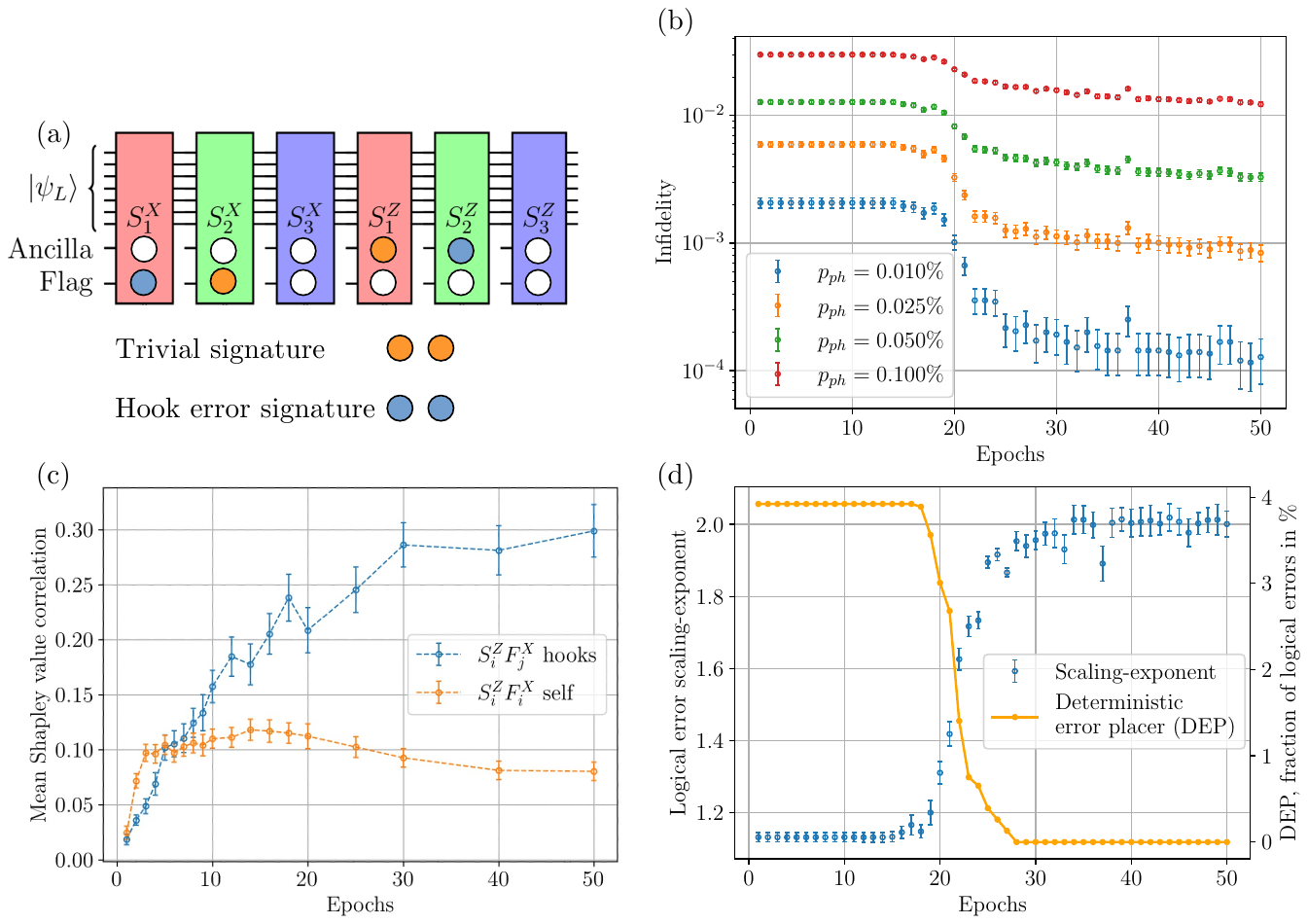}
    \caption{Monitoring the training of the RNN over 50 epochs. The learning of the handling of FT signatures (example in (a), blue) by the NN is shown. For that in (a) the two syndrome-flag pairs are illustrated. Here, the trivial (orange) combination corresponds to a weight-one data qubit error, while the hook error signature (blue) indicates a weight-two data qubit error.  In (b) the logical infidelity is displayed for different physical error rates. Sub-figure (c) shows the correlation of Shapley values of pairs of syndrome and flag bits that either indicate a hook error (blue) or are not relevant for FT (orange). According examples are given in (a). In (d) two figures of merit to quantify the FT of the decoder are shown. In blue, the scaling exponent of the logical error rate per round with the physical error rate is plotted. Further, a deterministic error-placer (DEP) is employed that iterates over all possible single faults in two rounds of QEC. For these instances, the orange curve shows the ratio of single faults that are uncorrectable for the RNN decoder.}
    \label{fig:monitoring}
\end{figure*}


It is an interesting question to ask how a NN learns certain competences during the training. Being able to answer this might provide insights that can help in optimizing the learning process or to stop it at a suitable point. In the context of fault tolerance in decoding, we aim to investigate this question in order to reliably determine when a NN decoder has learned to correct all single faults (for a distance-3 code). We analyze thereby the previously introduced RNN decoder and how it is learning to handle FT-breaking hook errors correctly in the Steane code with a flag-based readout scheme.

In order to do so, we approximate Shapley values for network instances during the training. As in the previous discussion, we aim to interpret the networks' capability to correct for hook errors by means of Shapley value correlations that align to  signatures of hook errors. Such Shapley correlations are now resolved over the training time.
Hence, it is of interest how and when the RNN decoder becomes fault-tolerant and whether we can deduce a good indicator of this learning transition in terms of the Shapley correlation.

To this end, the RNN model is saved after each training epoch; a training epoch being defined as the RNN having seen all training data once in a batchwise fashion. For each of these network instances and a verification data set, the Shapley values are calculated, and the decoder performance is tested.
We quantify this performance in different ways: First, in~\cref{fig:monitoring} (b) we show how the fidelity for different physical error rates behaves along the training. Further, we access in~\cref{fig:monitoring} (c) whether the RNN decoder training instances are FT. To this end, we show (i) the scaling exponent $a$ of the logical error rate per round against the physical error rate, i.e. $p_L\propto p_{ph}^a$, where a value close to $2$ would indicate FT as at least two faults are needed to induce a logical error. Further, (ii) the figure shows the fraction of logical faults occurring for the generation of all possible weight-one fault trajectories for two QEC cycles using a deterministic error placer (DEP). The DEP works by iterating over all possible circuit locations where a fault could occur, and placing the fault to evaluate whether the RNN decoder can recover the initial logical state or not.
Furthermore, we aim to quantify the evolution of how well the NN understands to consider the FT-breaking hook error signatures for decoding. We do this by means of the correlations between Shapley values as shown in~\cref{fig:monitoring} (d), where we compare two classes of Shapley value correlations: one corresponds to the hook error signatures, while the other is not related to the question of FT and serves as a baseline.

All of these figures of merit are evaluated for a single-output LSTM bit-flip decoder trained using $100.000$ training samples. Without decoding, a logical error occurs roughly for $4\%$ of the weight-one fault circuits in two QEC rounds as sampled with the DEP. Given this, one can understand the early stages of the training of the network. Here, as can be seen in \cref{fig:monitoring} (c), the failure rate of the RNN decoder is $4\%$ for the first $\sim$20 training epochs. While the network is on this plateau, it always assigns the trivial correction, which is not to perform any correction, independent of the syndrome. Note that also the infidelity in~\cref{fig:monitoring} (b) starts off with a plateau, which as well corresponds to the proportion of Monte Carlo samples which do lead to a logical error. Eventually, for the shown training process, the NN leaves the local loss-function minimum of assigning the trivial correction. 
At around epoch 16, the decoder starts to be able to decode the first few logical errors effectively, and it becomes fault-tolerant at epoch 27. At around the same time, the infidelity starts decreasing, but it keeps decreasing well beyond the point where the network achieves fault-tolerance. 

A corresponding evolution of the averaged hook error Shapley correlations for the bit-flip SRNN decoder is shown in~\cref{fig:monitoring} (d), where as a reference the average Shapley correlations of syndrome flag bit combinations, which do not correspond to a hook error, is shown. Interestingly, both curves start to deviate strongly when the FT performance indicators start to improve. It seems that all Shapley correlations increase first, although the ones that do not correspond to hook errors are quickly overtaken by the ones corresponding to bit-flip hook error configurations. After this break-even, they flatten out and start decreasing again. One can conjecture that the network initially decreases the loss function by attributing importance to incorrect flag-syndrome combinations, perhaps simply using the unspecific activation of both, syndrome and flag as a marker of logical errors. But then the NN quickly learns that only specific combinations are important for a good decoding. Altogether, the deviation of the two correlations can be used to track the starting point of the processes to learn the significance of the flag bit for the NN, yielding a first indicator for the start of the FT learning process.
Around epoch 30, the NN has managed to become FT as the logical error scaling exponent reaches a stable value of $2$. This FT decoding behavior is corroborated by the DEP test, which does not show any residual uncorrected weight-one faults. Accordingly, the Shapley value correlation of the bit-flip hook configurations flattens out above a value of $0.25$. This logistic behavior can be taken as a signal for the termination of the learning process of FT. The performance in terms of the logical error rate, however, continues to decrease further and one could even see minimal improvement beyond epoch 50. The reason for this is the learning of how to correct more weight-$2$ faults that occur. Despite this being an, in principle, desirable effect, one needs to carefully choose the point of termination of training as this learning of specific higher-weight faults, present in the training data, might be seen as over-fitting. For this reason, it is important to evaluate the network performance and properties for an independent data set as done in this analysis.


\section{Diagnosing NN malfunction for the two-headed NN}
\label{sec:Diagnosing NN malfunction for the two headed NN}

\begin{figure*}
    \centering
\includegraphics[width=1.5\columnwidth]{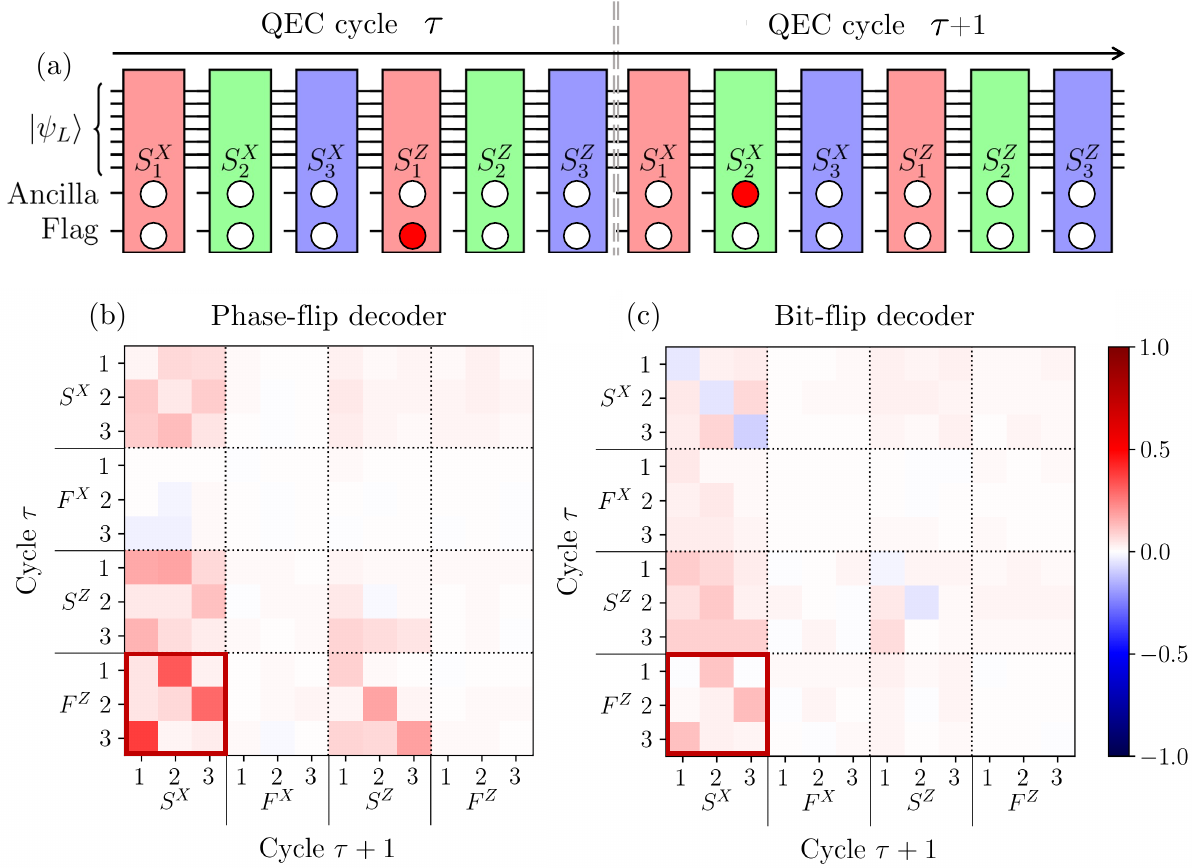}
    \caption{The Shapley-correlation analysis for the dual output RNN decoder, that is predicting simultaneously logical bit and phase flip parity. In (a) the typical syndrome flag signature for a $Z$-type hook error is shown, which stretches over two consecutive QEC cycles. In (b) the Shapley correlation for the logical phase flip decoding is shown, the quadrant of the expected syndrome flag correlation excess according to the hook signatures is highlighted. In (c) the Shapley value correlations between syndrome and flag bits of two consecutive rounds for the logical bit flip decoder are shown. An excess of Shapley value correlations in the same quadrant as for the phase flip decoder can be observed. These syndrome flag pair, however, carry no special information on the logical bit flip parity. In consequence, we can take these correlations as a sign for an incorrect use of phase-flip decoder logic to detect logical bit flips, pointing to a malfunction of the network.}
    \label{fig:dual-output-bit-flip-decoder-dt1}
\end{figure*}

The second Steane QEC decoder we will analyze in more detail in the following is based on a recurrent neural network that has two output neurons, each for predicting one of either logical bit or phase flip error parity. This network structure can be seen as an augmentation of the previous one, and it is also similar to the ones studied in Refs.~\cite{Baireuther2018,Baireuther2019}.  This means that both error types can be decoded within one forward-pass of the network, given the full syndrome flag volume as input. Using such a network, keeping the overall network size fixed, practically halves the computational cost of the decoder. However, the more important motivation for this proposed architecture is to harness the  correlation between $X$ and $Z$ syndrome more directly to potentially improve the decoding performance. The hope is that by predicting the logical $X$ and $Z$ parity in one pass, the network can more efficiently integrate these correlations in its decision-making process as both syndrome species are treated on the same footing in the training procedure. To underline this reasoning, one can think of the single output architecture; for it, half of the input information (e.g., the $Z$ syndrome to correct for $X$ errors) has far greater importance to compute the logical parity compared to the other part such that the less important part of the input might be pathologically underused.

For training the network, one must consider that labels of the desired decoder output are incomplete for each sample since for each execution of the memory experiment, the logical state can only be read out in the $X$ \textit{or} the $Z$ basis as it would be the case in a real-world experiment. This problem can be mitigated by using, alternatingly, only one of the output neurons for the backpropagation during training. In this setting, the cost function that is minimized adapts dynamically depending on the measurement basis to evaluate the correctness of the prediction of the appropriate output. Upon successfully training this new NN decoder, we will now investigate its performance and Shapley value correlations. In~\cref{fig:steane}, the logical error rate of the dual-output decoder can be compared to two single output LSTM networks. The single-output network variant outperforms the new architecture by quite a margin, reaching a pseudo-threshold at physical error rates roughly twice as high as for the dual-output decoder. We conjecture that there may be some unwanted internal network interference between the two computations of the respective logical parities, which causes the dual output LSTM network to underperform. This observation opposes the architectural ansatz to improve the decoding precision by having a dual output.

In the following, we analyze the decision-making of the dual-output NN decoder based on the Shapley interpretability method further.
Let us first consider the syndrome-flag signature of hook errors that would cause a logical phase flip. These generally span over two consecutive QEC cycles. An example of such a syndrome flag pair is illustrated in~\cref{fig:dual-output-bit-flip-decoder-dt1} (a) and rigorously all such signatures are given by the combinations $(F^t_{Z_i} S^{t+\Delta t}_{X_j})$ at $\Delta t=1$ for $j=i+1\mod 3$. Naturally, it is important for the phase-flip decoder to recognize these signatures.
In the correlation plots of~\cref{fig:dual-output-bit-flip-decoder-dt1} $(b)$ this can be seen as a correction excess when considering the dual RNN decoding of initial $\ket{\pm_L}$ states against logical phase flips. In sub-figure~\ref{fig:dual-output-bit-flip-decoder-dt1} $(c)$ the Shapley correlations for the bit-flip decoding with the dual RNN decoder is shown for syndrome-flag combinations of subsequent readout rounds, i.e. $\Delta t=1$. These correlations are not expected to show any excess, as no bit-flip hook error signatures exist here. On the contrary, the correlation excess of the phase-flip decoding in sub-figure,~\ref{fig:dual-output-bit-flip-decoder-dt1} $(c)$ can be observed as an artifact in the correlation structure of the bit-flip decoder in sub-figure~\ref{fig:dual-output-bit-flip-decoder-dt1} $(d)$.

This signature is an indicator of a badly tuned network where both species of syndrome and flag information influence the decoding decisions mutually. This mutual influence should in principle enable a better correction of correlated $X$ and $Z$ errors that originate from $Y$ errors. We observe the opposite, that the performance of respective bit- and phase-flip decoders is decreased, as shown in~\cref{fig:steane} (d). The understanding of this observation can now be aided by the previous discussion of Shapley correlations in the bit- and the phase-flip decoder. It seems that the RNN is not able to discriminate the respective information on bit- and phase-flip errors to a sufficient degree. Correspondingly, in this architecture, the phase-flip error-information can be seen as noise for the bit-flip decoding, rather than as additional information and vice versa.
It should however be noted that the correct hook error configurations are still most pronounced and that the dual-output network can still decode both syndrome bases fault-tolerantly. 
Altogether, this is a minimal example of the interpretation of a NN decoder where light could be shed on an internal malfunction that was noticeable as a sub-optimally performing decoder.
As a consequence, for the double-output architecture, we observe that this malfunction overshadows any positive effect of exploiting correlations of $X$ and $Z$ errors, which should potentially be more feasible for this NN architecture.


\section{Conclusion and Outlook}
\label{sec:Conclusion and Outlook}

In this work, we have employed an efficient local interpretation method for NNs, the Shapley value approximator DeepSHAP~\cite{chen2022explaining}. 
We have used it to explain the decoding decision of a RNN-based decoder for fault-tolerant operation of the Steane QEC code. The specific LSTM-based RNN architectures we analyze as examples are inspired by earlier works~\cite{Baireuther2019,Baireuther2018}. Our simulations confirm that these networks can be trained to become fault-tolerant decoders, and that they outperform sequential look-up-table decoders by a significant margin.
We derive an indicator, based on Shapley value correlations, of the NN for having learned a fault-tolerant decoding behavior. This understanding is based on the flag-fault-tolerant readout scheme that is used to measure the stabilizer generators of the Steane code. Evaluating appropriate Shapley value correlations is thereby independent of the performance analysis of the decoder. For simulated quantum memory experiments, the analysis of the scaling of the logical error rate with the physical error rate or the extensive placing of errors is a viable way to confirm FT. In an experimental setting, however, where this is not possible, the analysis of Shapley value correlations yields an alternative criterion.  

In our work, we further analyze the learning of correlations by the recurrent neural networks to determine a decoding decision. Most notably, we can identify all hook error signatures and correlations between $X$ and $Z$ errors. For future works, it can be interesting to consider other importance scores than the Shapley value which are suited more naturally to calculate feature-correlations, such as, e.g., the so-called `Shapley interaction values'~\cite{Rozemberczki2022, Dhamdhere2020shapley}. Lastly, we present a dual output LSTM network, decoding $X$ and $Z$ in parallel, which shows suboptimal performance after training. Employing our interpretability framework, this suboptimal performance can be understood.
We suggest this method as a tool to analyze shortcomings of NN decoders also in other settings to aid in the network-engineering.

In future work, it may be interesting to observe if and how neural-network decoders are able to handle noise models which include error types such as qubit loss, spatially correlated errors, biased noise, and non-Markovian noise. Moreover, it would be interesting to see how XAI explanation techniques can be used to optimize performance of NN-based decoders, both for the discussed Steane code but also for larger codes and more complex  FT protocols. Here, interpretable neural-network decoders could aid in the theoretical analysis of such larger codes by identifying and highlighting key features which neural networks were able to learn. It would be interesting to analyze the internal distribution of Shapley values in the NN to open the black box even wider.

\emph{Author contributions---}
LB and LK contributed equally to this project. LB conceived the project idea, LK set up the numerical tools and code. Both LB and LK performed the simulations, analyzed the data and prepared the figures. All authors contributed to the writing of the manuscript. LB and MM supervised the project.
\\

\emph{Data availability---}
The numerical code used in this work will be made available upon reasonable request.
\\

\emph{Acknowledgments---}
We acknowledge support by the Deutsche Forschungsgemeinschaft through Grant No.~449905436 and the ERC Starting Grant QNets through Grant Number 804247.
M.M.~furthermore acknowledges funding by the Deutsche Forschungsgemeinschaft (DFG, German Research Foundation) under Germany's Excellence Strategy – Cluster of Excellence Matter and Light for Quantum Computing (ML4Q) EXC 2004/1 – 390534769, funding by 
 by the Germany ministry of science and education (BMBF) via the VDI within the project NeuQuant (project number 13N17065), funding by the U.S. ARO Grant No. W911NF-21-1-0007, and from the European Union’s Horizon Europe research and innovation programme under grant agreement No 101114305 (“MILLENION-SGA1” EU Project). This research is also part of the Munich Quantum Valley (K-8), which is supported by the Bavarian state government with funds from the Hightech Agenda Bayern Plus. The authors gratefully acknowledge the computing time provided to them at the NHR Center NHR4CES at RWTH Aachen University (project number p0020074). This is funded by the Federal Ministry of Education and Research, and the state governments participating on the basis of the resolutions of the GWK for national high performance computing at universities.
\bibliography{references}

\widetext
\newpage
\begin{center}
\textbf{\large Supplemental Material for \textit{On the interpretability of neural-network decoders}}
\end{center}
\setcounter{equation}{0}
\setcounter{figure}{0}
\setcounter{table}{0}
\setcounter{section}{0}
\makeatletter
\renewcommand{\theequation}{S\arabic{equation}}
\renewcommand{\thefigure}{S\arabic{figure}}
\renewcommand{\thesection}{S\,\Roman{section}}
\renewcommand{\bibnumfmt}[1]{[S#1]}

\section{Neural network architecture, layout and training}
\label{sec:RNN_architecture}
In the following part, we summarize the technical details for network architecture and training for the  RNN decoders presented in the main text.
In order to efficiently approximate a target function $\hat f$ using neural networks to solve a problem such as decoding, their overall architectures have to be engineered carefully. The network layout such as number of layers, number of nodes per layer, and non-linear activation function, as well as the loss function and even the gradient descent algorithm all influence whether a network reaches an adequate approximation of the target function or not. Moreover, pre-processing of the training data or pre-training of the NN can also help the network learn more efficiently.
Please note that we do not perform an extensive optimization of hyperparameters, and we do not claim that the NNs considered in this work are optimal in any sense.

\subsection{Architecture}
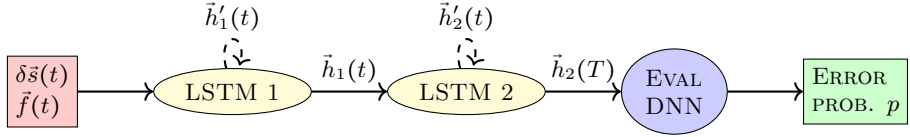
\begin{figure}[H]
\begin{center}
  \begin{tikzpicture}[]
    \tikzstyle{layer} = [draw=black, align=left]
    \tikzstyle{input} = [fill=red!20]
    \tikzstyle{output} = [fill=green!20]
    \tikzstyle{lstm} = [fill=yellow!20, ellipse]
    \tikzstyle{dnn} = [fill=blue!20, ellipse]
    \node[layer, input] (input) at (0, 0) {$\delta \vec s(t)$\\ $\vec f(t)$};
    \node[layer, lstm] (LSTM1) [right=1cm of input] {\sc LSTM 1};
    \node[layer, lstm] (LSTM2) [right=1cm of LSTM1]{\sc LSTM 2};
    \node[layer, dnn] (EVAL) [right=1cm of LSTM2] {\sc Eval\\ \sc DNN};
    \node[layer, output] (Output)[right=1cm of EVAL] {\sc Error\\ \sc prob. $p$};

    \draw[->, thick] (input) -- (LSTM1);
    \draw[->, thick] (LSTM1) -- node[above] {$\vec h_1(t)$} ++(LSTM2);
    \draw[->, thick] (LSTM2) -- node[above] {$\vec h_2(T)$} ++(EVAL);
    \draw[->, thick] (EVAL) -- (Output);

    \path (LSTM1) edge [loop above, dashed, thick] node {$\vec h'_1(t)$} (LSTM1);
    \path (LSTM2) edge [loop above, dashed, thick] node {$\vec h'_2(t)$} (LSTM2);
  \end{tikzpicture}
\end{center}
  \caption{Schematic illustration of the single output LSTM based RNN decoder. Depending on what labels the network is trained, it can either predict the logical bit- or phase flip probability. The left red box represents the time resolved syndrome and flag data, which is passed sequentially to two LSTM layers. A self-addressed arrow indicates their recurrent nature. The final state of the second LSTM layer, after $T$ inputs, is passed to a dense neural network (DNN) that maps its input to a single value $p$ that represents the decoding decision. }%
\label{fig:dual-lstm-singular}
\end{figure}

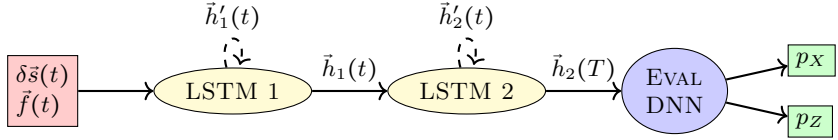
\begin{figure}[H]
\begin{center}
  \begin{tikzpicture}[]
    \tikzstyle{layer} = [draw=black, align=left]
    \tikzstyle{input} = [fill=red!20]
    \tikzstyle{output} = [fill=green!20]
    \tikzstyle{lstm} = [fill=yellow!20, ellipse]
    \tikzstyle{dnn} = [fill=blue!20, ellipse]
    \node[layer, input] (input) at (0, 0) {$\delta \vec s(t)$\\ $\vec f(t)$};
    \node[layer, lstm] (LSTM1) [right=1cm of input] {\sc LSTM 1};
    \node[layer, lstm] (LSTM2) [right=1cm of LSTM1]{\sc LSTM 2};
    \node[layer, dnn] (EVAL) [right=1cm of LSTM2] {\sc Eval\\ \sc DNN};
    \node[layer, output] (Output1)[right=1cm of EVAL.north east] {$p_X$};
    \node[layer, output] (Output2)[right=1cm of EVAL.south east] {$p_Z$};

    \draw[->, thick] (input) -- (LSTM1);
    \draw[->, thick] (LSTM1) -- node[above] {$\vec h_1(t)$} ++(LSTM2);
    \draw[->, thick] (LSTM2) -- node[above] {$\vec h_2(T)$} ++(EVAL);
    \draw[->, thick] (EVAL) -- (Output1);
    \draw[->, thick] (EVAL) -- (Output2);

    \path (LSTM1) edge [loop above, dashed, thick] node {$\vec h'_1(t)$} (LSTM1);
    \path (LSTM2) edge [loop above, dashed, thick] node {$\vec h'_2(t)$} (LSTM2);
  \end{tikzpicture}
\end{center}
  \caption{Schematic illustration of the dual output LSTM based RNN decoder, which can calculate the logical bit- and phase-flip error probability at once. The overall structure follows that of~\cref{fig:dual-lstm-singular}. It differs with respect to the final output, which here is desired to be the bit- and phase-flip parity. Accordingly, the dense neural network part has two output values $p_X$ and $p_Z$.}
\label{fig:dual-lstm-xz}
\end{figure}

The general structure of our RNN decoders, based on Refs.~\cite{Baireuther2018, Baireuther2019}, is visualized in~\cref{fig:dual-lstm-singular} and ~\cref{fig:dual-lstm-xz} for the single and the dual output RNN, respectively. This structure is based on having two LSTM layers followed by a dense evaluation network that is fed with the final LSTM state of the second LSTM layer.
The detailed architectural NN parameters are given in~\cref{tab:NN_architecture} for the two RNN decoders that are presented in the main text. The dropout layers that are included are only active during the training, as it is usually the case.

\begin{table}[H]

\begin{center}
\begin{tabular}{ccccc}
  \toprule
  Layer & Layer Type & Neurons & Activation Function & Note \\
  \midrule
  \midrule
  1     & Masking    & 0       & -\\
  2     & LSTM       & 36      & Sigmoid\textsuperscript{(1)} / ReLU\textsuperscript{(2)} & Sequential output\\
  3     & LSTM       & 36      & Sigmoid\textsuperscript{(1)} / ReLU\textsuperscript{(2)}& Single output\\
  4     & Dense      & 48      & ReLU\\
  5     & Dropout    & 0       & - & Training only, $p=0.2$\\
  6     & Dense      & 24      & ReLU\\
  7     & Dropout    & 0       & - & Training only, $p=0.2$\\
  8     & Dense      & 12      & ReLU\\
  8     & Dropout    & 0       & - & Training only, $p=0.2$\\
  9     & Dense      & 1 / 2\textsuperscript{(3)}   & Sigmoid\\
  \toprule
\end{tabular}

  \textsuperscript{(1)}: Gated Interactions
  \textsuperscript{(2)}: Output gate
  \textsuperscript{(3)}: 2 outputs for the dual decoder
\end{center}
\caption{Summary of the used network architectures, specifying the width and arrangement of each layer. The single-valued RNN model has one output neuron in the final layer, while the $X+Z$ decoder has two outputs. For the dropout layer $p=0.2$ refers to a $20\%$ probability for each neuron to be excluded in a training step.}
\label{tab:NN_architecture}
\end{table}

\subsection{Network optimization}
One can think of a network, parametrized by weights $\vb W$, as an ansatz spanning a space of possible functions $\mathcal F$. In order for the network to approximate a target function $\hat f: \vec x \mapsto \vec y$, the weights $\vb W$ must be carefully chosen or optimized.

In supervised learning, the weights are learned from a set of samples $\{\vec x_i, \vec y_i\}_i$. The training samples are assumed to be related to each other by some unknown target function $f$ such that $\vec y_i = f(\vec x_i)$.
In order to solve this optimization problem, one first defines a measure of distance between functions in the function space $\mathcal F$. This measure may then be minimized using a gradient descent algorithm to find the optimal function $f$ which approximates $\hat f$. The measure of distance of functions is referred to as the \textit{loss function} $\mathcal L$.  The weights are updated by taking a step in the negative direction of the gradient. The update rules for the individual weights are then propagated through the network in reverse order using the back-propagation algorithm~\cite{BackpropAlgo}, 
\begin{equation}
  \vb W^{(\tau+1)} = \vb W^{(\tau)} - \gamma \nabla_{\vb W} \mathcal L(\vb W^{(\tau)}).
\end{equation}
Here, $\gamma$ is the learning rate that can be considered as a hyperparameter which tunes the size of steps and thereby the speed of learning. Increasing $\gamma$ makes the network learn faster, but setting it too high may jeopardize its convergence, as the loss function may jump around local minima. It is a common practice in supervised NN training to process the training data in batches of fixed size and update the weights based on an average of each batch. This method is also referred to as stochastic gradient descent, as not all training data is used to approximate a gradient as well as possible. The random sampling of data-subsets (batches) causes the gradient to vary stochastically between every weight update. This stochasticity as shown to improve the convergence of NNs for generic problems, as it helps to escape local minima~\cite{bottou2007tradeoffs}.

To train the NNs employed in the main text, we further use a so-called optimizer by which the learning rate becomes non-constant and dependent on the previous learning process. In particular, we choose the Adam (Adapted Moment) optimizer out of many popular implementations in the Tensorflow/Keras library~\cite{tensorflow}. It is a stochastic gradient descent optimizer based on the adaptive estimation of low-order moments~\cite{kingma2017adam}. As the name implies, Adam performs gradient descent that is additionally influenced by  the mean $m_\tau$ and variance $v_\tau$ of the past gradients~\cite{kingma2017adam}. Then, a moving average of these quantities is used to adapt the effective learning rate as
\begin{equation}
  w^{(\tau+1)} = w^{(\tau)} - \gamma \frac{\hat m_\tau}{\sqrt{\hat v_\tau + \epsilon}},
\end{equation}
with
\begin{equation}
   m_\tau = \frac{ \beta_1 \hat m_{\tau-1} + (1-\beta_1) 
  \nabla_{w} \mathcal L\left(w^{(\tau)}\right)}%
  {1-(\beta_1)^{\tau}},
\end{equation}
\begin{equation}
   v_\tau = \frac{\beta_2 \hat v_{\tau-1} + (1-\beta_1) 
  \left(\nabla_w \mathcal L\left(w^{(\tau)}\right)\right)^2}%
  {1-(\beta_2)^\tau}.
\end{equation}
Here $w^{(\tau)}$ is a specific weight in $\vb W^{(\tau)}$, $\beta_1$ and $\beta_2$ are hyperparameters which tune the update rate of the moving averages of the mean $ m_\tau$ and the second moment $ v_\tau$ respectively with $\beta_i\in[0, 1)$.
Further, $\hat m_\tau$ denotes the average $m_\tau$ with respect to the batch-data as well as $\hat v_\tau$ is the averaged and centralized $v_\tau$.
The $\epsilon$-hyperparameter is set to some small, finite value, which guarantees numerical stability. The typical values for the hyperparameters, which we also employ are $\beta_1=0.9$, $\beta_2=0.999$, and $\epsilon=1\cdot10^{-7}$~\cite{kingma2017adam}. These are the default values in the Tensorfow/Keras implementation of Adam.

Training is further divided into epochs. In one epoch, the network is shown each of the training samples exactly once. Training the network for multiple epochs tends to increase the network's accuracy. Here, with accuracy we mean the tested performance of the NN on a data set that is never used for the NN optimization.
At some point during the training, the network may converge in the sense that adding more epochs may reduce the accuracy with the reference data. For large numbers of epochs, the network may even start \textit{overfitting} the data. Here, the network learns to perfectly match the training data but starts performing worse on the data it has not yet seen. For this reason, an independent set of data not used during training is used to validate the network's performance. If the validation accuracy saturates or starts decreasing, the training process may be stopped.

\subsection{Loss function}
For binary classification tasks, the Binary Cross Entropy (BCE) is a suitable loss function $\mathcal L$~\cite{Liu2017}.
It is used for the training single output RNN decoder that is presented in the main text.
The BCE loss is defined as
\begin{equation}
  \mathcal L(p, q) = - (p \log(q) + (1-p)\log(1-q)), \quad p\in\{0, 1\},q\in(0, 1),\\ 
\end{equation}
where $p=y$ is the target label and $q=f(\vec x)$ is the model prediction. The label $y\in\{0, 1\}$ represents the class. Note that if $y=0$, the loss function reduces to $\mathcal L = -\log (1-q)$, and if $y=1$ then $\mathcal L = -\log q$. As $q\to p$, the binary cross entropy monotonically decreases, which means that by adjusting the weights to minimize the loss function, the network learns to modify its classification towards the correct label. For binary classification problems, it also tends to be useful to choose an activation function for the output layer which limits the model's output to some finite domain. Commonly, the output layer uses a sigmoid activation, limiting the final output to the $(0, 1)$ domain. In this sense, the model output can be interpreted as the probability that an input $\vec x_i$ belongs to a specific class.

For training the dual-output RNN that shall perform a decoding for the logical bit- and phase-flip parity simultaneously, the loss function needs to be adapted as for every run of the quantum memory protocol, there is only a single training label for either of the $X$ or $Z$ basis  available while the other logical parity is inaccessible. This augmentation can be done by defining a masked binary cross entropy, where the unknown logical parity bit is replaced by a mask $m$. The target vector for the training is then either $(p_X, m)$ or $(p_Z, m)$ and one can define
\begin{equation}
  \mathcal L = \begin{cases}
    p_X\log (q_X) - (1-p_X)\log (1- q_X), & p_Z=m\\
    p_Z\log (q_Z) - (1-p_Z)\log (1- q_Z), &  p_X=m\\
  \end{cases}
\end{equation}
as the loss function.
The training with this loss function implies that for each individual back-propagation, only one of the outputs serves as the starting point. As a consequence, the weights, involved in the connection of the last dense NN layer to the two output nodes, are only updated every second time on average. All other weights are updated always, but if the $X$ and $Z$ decoding are understood as independent problems, there might be a competition among the network parameters to solve either of them. The idea of the dual-output RNN construction is that this competition is resolved internally during the training and that the joint solving of $X$ and $Z$ can be beneficial for both as they are not entirely independent problems due to the presence of $Y$ errors that affect both.

\section{Recurrent neural network architecture}
\label{sec:RNN}
In the following, we provide details on the recurrent nature of the NN, that are used in the main text and provide technical details of the used recurrent unit---the long short-term memory (LSTM) cell.

\subsection{Recurrent neural networks}
Recurrent neural networks (RNNs) are a type of neural network where neurons may loop back onto themselves~\cite{Schmidt2019} such that the output of one layer may be reprocessed in the same layer together with a subsequent input at another time. In this sense, recurrent neural networks have both a spatial layer-to-layer interaction and a temporal intra-layer interaction. This makes RNNs especially capable of learning temporally correlated features in data. Furthermore, RNNs are inherently capable of processing sequences of data, even sequences of variable length.
RNNs have been applied to, amongst others, language modelling and generating text~\cite{Schmidt2019, Wang2015, Chen2017}, speech recognition~\cite{Schmidt2019, Sak2014}, generating image descriptions and video tagging~\cite{Schmidt2019}. QEC decoders for quantum memories, where encoded quantum information is required to be stored for variable times, is also a natural candidate problem to be solved by RNNs~\cite{Baireuther2018,Baireuther2019}. Furthermore, RNNs have been proven to be Turing-complete~\cite{RNNTuringComplete}.
This makes RNNs an attractive architecture for learning complex algorithmic decoding schemes.
A drawback of RNNs is that they suffer from the vanishing gradient problem~\cite{Schmidt2019}, since the networks can reach exceptionally deep temporal depths. For this reason, training RNNs can also be computationally very costly~\cite{RNNVanishingGradient}. These problems may in part be circumvented using specialized RNN unit architectures like the LSTM unit, which is defined rigorously in~\cref{sub:LSTM-cell}.

\subsection{The Long Short-Term Memory Cell}%
\label{sub:LSTM-cell}
The Long-Short-Term Memory (LSTM) unit is a type of RNN layer designed to learn long-term correlations in training data~\cite{Hochreiter2001}. The term long short-term memory refers to the fact that the LSTM unit has a relatively \textit{long} short-term memory. LSTM units work by using a dedicated memory `cell' value which, along with the unit's outputs, is also passed to the same unit at the next time step. This dedicated memory state allows these networks to typically retain information over extended periods of well over 1000 time steps~\cite{Schmidt2019}. Note that the terms LSTM unit and LSTM layer are used interchangeably. Each unit or layer has a fixed size that specifies the size of the input and output vector as well as the cell state vector.

\textit{Computational structure of LSTMs---}Each LSTM unit contains 3 subroutines, so-called gates: input, forget, and output. These gates control the flow of information into and out of the memory cell. The input gate regulates the flow of new information into the memory cell, the forget gate decides what information to discard, and the output gate determines what information to output based on the current cell state. Each of the gates is implemented as a miniature neural network with own weights that are learned. The purpose of the LSTM unit with its subroutines is to calculate a new output $\vec H_{t}$ and cell state $\vec C_{t}$ from an input $\vec X_t$ and the previous output $\vec H_{t-1}$ and previous cell (memory) state $\vec C_{t-1}$.
In our framework, the input to a LSTM unit/layer is either the initial syndrome and flag input or the output of the spatially previous LSTM layer.

Let's consider first the updating of the cell state $\vec C_{t-1}\rightarrow \vec C_{t}$, where the first gated interaction is the forget gate. It is given by the current input $\vec X_t$, previous output $\vec H_{t-1}$ and is parametrized by $\{\vb W_{xf},\vb W_{hf}, \vec b_f\}$
\begin{equation}\label{eq:forget}
  \vec F_t = \sigma(\vb W_{xf}\vec X_t + \vb W_{hf}\vec H_{t-1} + \vec b_f),
\end{equation}
where $\sigma$ is the gate-wise sigmoid function, such that the output values of this gate range between $(0, 1)$. The length of the forget-vector $\vec F_t$ equals the size of the LSTM layer as well as the output or the cell-state vector $\vec H_{t-1},\vec C_{t-1}$. The corresponding weight matrices are square, while the weight matrix of the input vector $\vb W_{xf}$ does not have to be square.
This forget-vector $\vec F_t$ is now multiplied pointwise with the previous cell state, causing some part of the cell state to be `forgotten'. This element-wise multiplication will be denoted as $\vec C_{t-1}\odot \vec F_t$.
To update the cell state, we further use the so-called input gate. It performs the information update and can be broken up into the selection of a new candidate memory ${\vec C'}_t$ and a filter ${\vec I'}_t$ which lets only a part of the candidate memory through:
\begin{equation}\label{eq:input-filter}
  {\vec I'}_t = \sigma(\vb W_{xi}\vec X_t + \vb W_{hi}\vec H_{t-1} + \vec b_i),
\end{equation}
\begin{equation}\label{eq:candidate}
  {\vec C'}_t = \tanh(\vb W_{xc}\vec X_t + \vb W_{hc}\vec H_{t-1} + \vec b_c),
\end{equation}
\begin{equation}\label{eq:input}
  \vec I_t = {\vec C'}_t \odot {\vec I'}_t.
\end{equation}
Here, the $\tanh$ function is also to be understood as an element-wise application. Note that four new weight matrices and two new bias vectors are introduced that can be trained independently.
The new memory state $\vec C_t$ is then given by the forget and input gate interaction as
\begin{equation}
  \vec C_t = \vec C_{t-1}\odot \vec F_t + \vec I_t.
\end{equation}

Lastly, the output state $\vec H_t$ is calculated by means of the output gate and the updated cell state. The output gate, again, depends on the current input and the last output vector and is equipped with own weight and bias parameters,
\begin{equation}
  \vec O_t = \sigma(\vb W_{xo} \vec X_t +\vb W_{ho} \vec H_{t-1} + \vec b_o).
\end{equation}
To calculate the output state, each $\vec O_t$ value is weighted by the updated cell state that is regularized by a $\tanh$ function,
\begin{equation}
  \vec H_t = \vec O_t \odot \tanh(\vec C_t).
\end{equation}
This output becomes the input of the same LSTM unit at the next time-step  as well as being passed to the next LSTM unit. 

\section{The Shapley Value}%
\label{sec:shapley_definition_properties}
The Shapley value may be used for feature importance attribution in machine learning by mapping it to a `feature selection game'~\cite{Rozemberczki2022}. In this section, we provide some mathematical properties of the exact Shapley value as defined in Def. 3 in the main text, as well as a more detailed explanation on the approximative numerical method we employ to calculate it. In doing so, we will also comment on the software tools we utilize and how they work.

\subsection{Properties}
The Shapley value has the following desirable qualities as a credit assignment method, including fairness, efficiency, and symmetry. We will outline these properties following~\cite{Rozemberczki2022}. Hereby, efficiency refers to the evaluation of an explanation model, not the complexity of obtaining it. These qualities can be formalized using  the \textit{null player}, \textit{efficiency}, \textit{symmetry}, and \textit{linearity} properties. We will now formally introduce these concepts and give context as to why they are relevant in the context of XAI.

\begin{definition}[Null player]
  A player $i$ is a null player if $v(S\cup\{i\})=v(S)\ \forall\ \mathcal{S}\subseteq\mathcal{N}\setminus\{i\}$. A solution concept satisfies the null player property if for every solution vector and every null player $i$ it holds that $\phi_i(\mathcal{N}, v)=0$.
\end{definition}
This means that a solution concept that satisfies the null player property never assigns value to a player that does not influence a given result. This is desirable in XAI as irrelevant features should not be assigned credit.

\begin{definition}[Efficiency]\label{def:efficiency}
  A solution concept is efficient if, for every game and every solution vector, it holds that $\sum_i \phi_i(\mathcal{N}, v)=v(\mathcal{N})$. 
\end{definition}
A solution concept that satisfies the efficiency property satisfies that all solution vector entries sum to the output of the model as a whole. In XAI, this property allows for a quantitative interpretation of the contribution scores~\cite{Rozemberczki2022}. 

\begin{definition}[Symmetry]
  A solution concept satisfies the symmetry property if for all symmetric players $i,j\in\mathcal{N}$ $\phi_i(\mathcal{N}, v)=\phi_j(\mathcal{N}, v)$ where a symmetric player $i, j$ is defined as, $v(\mathcal{S}\cup \{i\})=v(\mathcal{S}\cup\{j\})\ \forall\ \mathcal{S}\subseteq\mathcal{N}\setminus\{i,j\}$
\end{definition}
The symmetry property ensures that two features with the same marginal contribution are assigned the same relevance. This ensures that relevance is fairly distributed over all inputs in the machine learning model.

\begin{definition}[Linearity]
  A single-valued solution concept satisfies linearity if for any two games $(\mathcal{N}, v)$, $(\mathcal{N}, w)$ and for the solution vector of the game $(\mathcal{N}, v + w)$
  \begin{equation*}
    \phi_i(\mathcal{N}, v + w) = \phi_i(\mathcal{N}, v) + \phi_i(\mathcal{N}, w)
  \end{equation*}
\end{definition}
Rozemberczki \textit{et al.}~\cite{Rozemberczki2022} give an example of how this property can be advantageous in the context of binary classifiers operating on independent datasets, however this is not relevant for our discussion.

It was shown that \textit{any} single-valued solution concept that satisfies all these properties is uniquely defined by the Shapley value~\cite{Shapley1953, Rozemberczki2022}. This means that a solution concept that satisfies all these properties can only be the Shapley value, making the Shapley value well suited to feature attribution in machine learning models from a theoretical perspective.~

\subsection{Shapley value approximation}
\label{subsec:shap_approx_theory}
First, we define the \textit{universal explainability game} as a candidate for a model-agnostic feature attribution XAI method~\cite{Rozemberczki2022}.
\begin{definition}[Universal explainability game]
  Let $f(\cdot)$ denote a machine learning model of interest and let the set of players be the features of a single data instance $\mathcal{N}=\{x_i\ |\ i\in\{1, \ldots, n\}\}$. The characteristic function of a coalition $\mathcal{S}\subseteq \mathcal{N}$ is $v(\mathcal S)=f(\mathcal S)$ calculated from the subset of feature values.
\end{definition}

A caveat regarding the application of the Shapley value to ML models is that ML models tend to only be able to handle fixed-sized inputs, which means that turning off features is impossible or would require retraining of the model. As repeated retraining is computationally intensive and often undesirable, feature exclusion is usually approximately implemented by replacing a value of a specific feature in a specific input sample with the mean of that feature over a large dataset~\cite{LundbergLee2017, Rozemberczki2022}. Formally, one can define the function $f_\mathcal{S}$ which approximately includes the features $\mathcal S$ and excludes all other features in $\mathcal N$ as follows~\cite{LundbergLee2017}
\begin{equation}\label{eq:feature-exclusion}
    f_{\mathcal S}(\vec x) = E[f(X_i) | X_i = x_i\in \vec x\, \forall\, i\in\mathcal S].
\end{equation}
Here $X_i$ is considered a random variable as it corresponds to one input sample drawn from the distribution of input variables. The notation $X_i=x_i$ means that the random variable is fixed to a specific value $x_i$, and the features $i\in \mathcal S$ are included features.
Apart from the question of feature exclusion and retraining, which would correspond to a constant computational overhead, a straight-forward computation of the Shapley value entails the computation of exponentially many contributions of subsets of input features.
More precisely, the exact calculation of the Shapley value comes with a time complexity of $O(2^{\abs{\mathcal N}})$. When evaluating large datasets with many features, it is therefore essential to come up with approximations to the Shapley value that reduce this computational cost. There are a number of state-of-the-art approximations for universal explainability, including linear regression approaches~\cite{LundbergLee2017, frye2021shapley, covert2021improving} and Monte Carlo sampling~\cite{yuan2021explainability} all with $O(\abs{\mathcal N})$ time complexity.  

In this work, we use a neural network specific method called DeepSHAP introduced by Lundberg \textit{et al.}~\cite{LundbergLee2017} which is included as part of the open source SHAP library~\cite{SHAPGit}.
To explain the entailed approximations and the working principle, we assume that the characteristic function $f$ of the black box model that we want to interpret using the Shapley value, is linear:
\begin{equation}\label{eq:linear-function}
  f(\vec x) = \beta_0 + \beta_1 x_1 + \cdots + \beta_n x_n.
\end{equation}
With this, the calculation of the Shapley value can be greatly simplified.
Furthermore, we assume that feature exclusion is defined as in
Equation~\ref{eq:feature-exclusion}. A function with features $\mathcal S\subseteq \mathcal N$ included is then given by
\begin{equation}
  f_\mathcal{S}(\vec x) = \beta_0 + \sum_{i\in \mathcal S}\beta_i x_i + \sum_{j \in \mathcal N\setminus \mathcal S}\beta_j E[x_j].
\end{equation}
Note now that the marginal contribution $\Delta f_i = f_{\mathcal S \cup i}-f_{\mathcal S\setminus i}$ reduces to
\begin{equation}\label{eq:linear-marginal}
  \begin{split}
    \Delta f_i(\vec x) =&\ \beta_0 + \sum_{k\in{S\cup i}}\beta_k x_k + \sum_{l \in \mathcal N\setminus (\mathcal S\cup i)}\beta_l E[x_l] \\
    &\ - \left(\beta_0 + \sum_{k\in{S\setminus i}}\beta_k x_k + \sum_{l\in\mathcal{N}\setminus(\mathcal S\setminus i)}\beta_l E[x_l]
  \right)\\
    =&\ \beta_i (x_i - E[x_i]).
  \end{split}
\end{equation}
When plugging the result of~\cref{eq:linear-marginal} into the definition of the Shapley value (Def. 3 of the main text) one can derive that the Shapley values are given by
\begin{equation}
  \phi_i = \beta_i (x_i - E[x_i]),\quad \phi_0 = E[f(X)].
  \label{eq;shap_lin_approx}
\end{equation}
where $E[f(X)]$ is the mean, according to the data distribution also called baseline output of the model. Note that this also implicitly assumes feature independence, i.e. $x_i\neq x_i(x_1, x_2, \ldots, x_n)$.
As a result of this approximation, the entire combinatorics problem is removed, making the calculation of Shapley values vastly less computational costly with a time complexity of $O(\abs{\mathcal{N}})$.

The assumption that the model is linear might seem a crude approximation, but by linearly approximating the model for individual inputs using the neural network's internal structure, good linear approximations around specific inputs may be found. To obtain said linear approximations of NN models, backpropagation can be harnessed. 
The approximated Shapley values may be interpreted as exact Shapley value solutions to an approximated model.
An implementation of this is given by DeepSHAP, a Shapley value approximation method for neural networks, which is a method based on the DeepLIFT XAI method that is introduced in the following.

\subsection{Numerical implementation of the Shapley value approximation}
\label{subsec:num_shap_approx}
\textit{DeepLIFT}\textit{---} is a local XAI method that backpropagates a relevance score for an input relative to some reference input. Such relevance score is meant as a value attributed to every neuron depending on the overall state of the NN.
We explain the working of DeepLIFT following~\cite{DeepLIFT2017}, starting by defining some NN-related concepts and quantities.
\begin{definition}[Input and target neuron activations]
  Let $t$ be some target output neuron and let $\vec x$ be an input (possibly in some intermediate layer or set of layers) necessary and sufficient to compute $t$.
\end{definition}
\begin{definition}[References]
  Let $\vec x^0$ be some `reference input'. Suppose activation of the target neuron with inputs $x_i$ is defined as $t = f(\vec x)$, then the reference activation of the target neuron is defined to be
  \begin{equation}
    t^0 = f(\vec x^0).
  \end{equation}
\end{definition}
\begin{definition}[Delta from reference]
   Define $\Delta x_i = x_i - x_i^0$ and $\Delta t=t-t^0$ to be the delta from reference (DFR) activation for the input and target, respectively.
\end{definition}
DeepLIFT backpropagates a relevance score based on DFR activations. Therefore, it assigns a relevance score $R_{\Delta x_i \Delta t}$ to $\Delta x_i$ for its contribution in producing the output $\Delta t$. At all times, the backpropagated relevance is thereby conserved by the \textit{summation to delta property}.
\begin{definition}[Summation to delta property]\label{def:deeplift-sum-to-delta}
  The contributions $R_{\Delta x_i \Delta t}$ by definition conform to the following summation to delta property
  \begin{equation}
    \sum_{x_i\in\{x_i\}_i} R_{\Delta x_i \Delta t} = \Delta t.
  \end{equation}
\end{definition}
For well-behaved activation functions, the output is locally linear in its inputs~\cite{DeepLIFT2017}. This motivates the concept of \textit{multipliers}, which define how the relevance score is backpropagated.
\begin{definition}[Multipliers]\label{def:deeplift-multipliers}
  For a given neuron $x$ with DFR $\Delta x$ and a target neuron $t$ with DFR $\Delta t$ we wish to compute the relevance $R_{\Delta x \Delta t}$. To this end the multiplier $m_{\Delta x \Delta t}$ is defined to be
  \begin{equation}
    m_{\Delta x \Delta t} = \frac{R_{\Delta x \Delta t}}{\Delta x}.
  \end{equation}
\end{definition}
In other words, the multiplier is the contribution of $\Delta x$ to $\Delta t$ divided by $\Delta x$. The multipliers are used to backpropagate the relevance. There exists a close analogy between multipliers and the partial derivative $\frac{\partial t}{\partial x}$, except that the multipliers use finite differences.


\begin{definition}[Linear Rule for relevance assignment]
For a linear network the Linear Rule is an exact rule of defining relevance scores. The relevance of $\Delta x_i$ leading to an output $\Delta y$ is defined to be
  \begin{equation}
      R_{\Delta x_i \Delta y} = w_i\Delta x_i.
  \end{equation}
\end{definition}

From this definition, the relevance score and multipliers can be defined for every neuron. By defining multipliers recursively using the linear rule, relevance scores can be backpropagated. By using the \textit{chain rule for multipliers}, a direct multiplier between the output and input domains can be defined.

\begin{definition}[Chain rule for multipliers]
  Suppose we are given the inputs $\{x_1,\ldots, x_n\}$ with hidden neurons $\{y_0,\ldots, y_m\}$ and some target output $t$. Given the values for $m_{\Delta x_i \Delta y_j}$ and $m_{\Delta y_j\Delta t}$ the following definition is consistent with Definition~\ref{def:deeplift-sum-to-delta}.
  \begin{equation}
    m_{\Delta x_i \Delta t} = \sum_j m_{\Delta x_i \Delta y_j} m_{\Delta y_j \Delta t}
  \end{equation}
\end{definition}

The choice of reference is heuristic in the DeepLIFT framework. When setting the reference of all inputs and activations to zero, DeepLIFT becomes equivalent to layer-wise relevance propagation in networks with ReLU activations and semi-negative bias constraints, and can be seen as an approximation to calculating local derivatives~\cite{DeepLIFT2017, LundbergLee2017}, see also~\cref{sec:LRP} for the definition of layer-wise relevance propagation.

\textit{DeepLIFT for approximating the Shapley Value---} Following~\cite{LundbergLee2017} and based on the findings in~\cref{subsec:shap_approx_theory}, one can use the DeepLIFT method to approximate Shapley values. By using the linear DeepLIFT rule, one can define the reference input as the expectation values of inputs $E[x]$ according to the distribution of the training data. In doing so, one obtains an efficient approximation to the Shapley values for deep networks. DeepLIFT linearizes the network locally, and by fixing the reference according to~\cref{eq;shap_lin_approx}, the output contributions may be interpreted as Shapley values. Including or excluding a feature, as discussed for the computation of Shapley values, can be thought of as using the actual value (including) versus using the reference value (excluding). The reference values are computed as a sample average of the dataset $\{x_i\}_i$.
Given a network with input values $\vec x=\vec x^{(0)}$, intermediate layer activations $\vec x^{(l)}$, and output values, $f(\vec x)=\vec x^{(L)}$ we define the Shapley values on the level of the input as $\phi_i^{(0)}=\phi_i^{(0)}(f, \vec x)$ and set the Shapley values of the output layer to $\phi^{(L)}=f(\vec x)$. One can then define the multipliers to recursively backpropagate Shapley values between layers
\begin{equation}
  \begin{split}
    m_{x^{(l)}_j, f_k} &= \frac{\phi^{(l+1)}_k}{x_j^{(l)} - E[x_j^{(l)}]}, \quad \phi^{(l+1)}_k=\phi(f_k, \vec x^{(l)}). \\
  \end{split}
\end{equation}
Using the chain rule, one can define the effective multiplier from output to input domain first and then calculate the Shapley value in one step
\begin{equation}\label{eq:DeepSHAP-approx}
  \begin{split}
    m_{x^{(0)}_i, f^{(L)}} &= \sum_{j,k,...,q}m_{x^{(0)}_i f^{(1)}_j} m_{x^{(1)}_j f^{(2)}_k} \ldots m_{x^{(L-1)}_q f^{(L)}}\\
    \phi^{(0)}_i &\approx m_{x^{(0)}_i f^{(L)}}\left(x^{(0)}_i - E\left[x_i^{(0)}\right]\right)\quad \text{(Linearization)}.
  \end{split}
\end{equation}

\textit{SHAP library---} The SHAP Python library~\cite{SHAPGit} offers an out-of-the-box implementation of the DeepSHAP algorithm, which works with Tensorflow/Keras or Pytorch models. The SHAP library uses the inbuilt Tensorflow graph-based gradient calculator for an efficient implementation. It calculates the Shapley values by the following steps:

\begin{enumerate}
  \item Provide a trained neural network model $f$ implemented in Tensorflow.
  \item Define a background dataset $\vb B = (\vec b_1, \vec b_2, \ldots, \vec b_N)$ with N background (input) samples $\vec b_i$, generally speaking chosen at random from an independent dataset from the data points we want to explain. This way, there are no systematic correlations between the chosen background distribution and input samples.
  \item Provide an input sample $\vec x$ for the calculation of its respective Shapley values.
  \item Shapley values $\vec \phi$ are calculated by the following equation (in vector notation):
    \begin{equation}
      \vec \phi = 
      \underbrace{(\tilde \nabla f^{-1}(f(\vec x)))}_{\vec m_{\vec x f(\vec x)}} \odot 
      \underbrace{\text{mean}(\vec x \ominus \vb B )}_{\vec x - E[\vec x]}
      \quad 
    \end{equation}
    Here, $\tilde \nabla f^{-1}(f(\vec x))$ means a modified backwards gradient implementing DeepLIFT's backpropagation from output to the input for the model $f$. This is implemented in Tensorflow~\cite{tensorflow}. $\vec x \ominus \vb B$ means the input vector minus the matrix of all background samples $\vb B$, where the output is a matrix of the same shape as $\vb B$, and the mean is then taken along the background subtracted samples. The first and second part are then multiplied element-wise to obtain the Shapley values.
\end{enumerate}

The DeepSHAP method is able to utilize the Tensorflow/Keras gradient implementation. Consequently, it is applicable in the framework for all NN types that are implementable and trainable with Tensorflow/Keras. This includes the recurrent LSTM networks used in this work.
Compared to the layer-wise relevance propagation methods presented in~\cref{sec:LRP}, there is no need for heuristic choices to define how the backpropagation is conducted. It is conducted as explained above to approximate the Shapley value as a relevance score, see~\cref{subsec:shap_approx_theory}. This makes it a solid candidate for XAI explanations of neural networks in general, and is used for the interpretation analysis of the RNN decoders in the main text.

\subsection{Assessing the validity of the approximation}
\begin{figure}[t]
\begin{center}
  \begin{subfigure}[]{0.40\textwidth}
  \begin{center}
    \includegraphics[trim=45 00 45 00, clip, width=\textwidth]{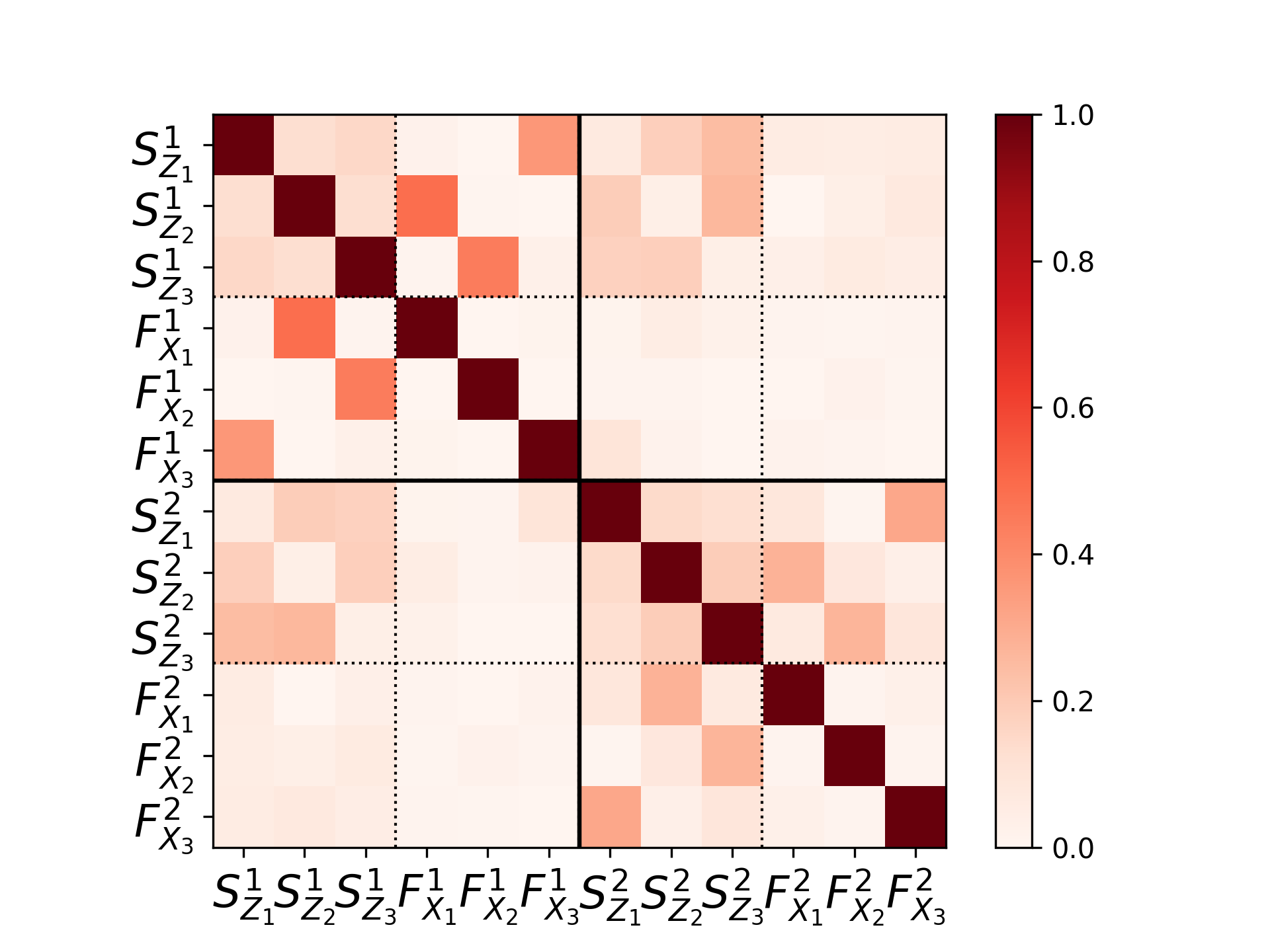}
  \end{center}
  \caption{}
  \label{fig:min-shap-exact}
  \end{subfigure}
  \hfill
  \begin{subfigure}[]{0.40\textwidth}
  \begin{center}
    \includegraphics[trim=45 00 45 00, clip, width=\textwidth]{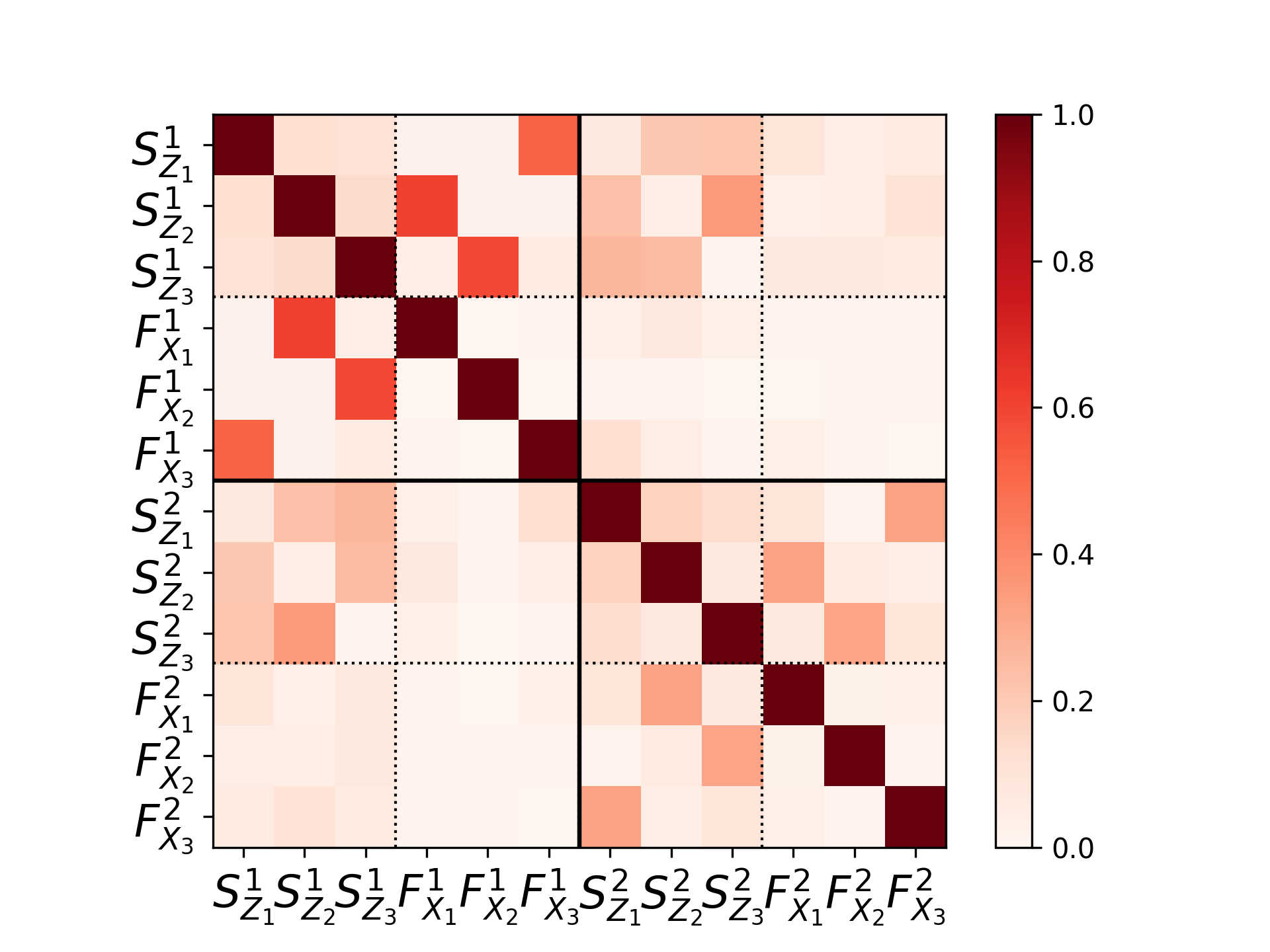}
  \end{center}
  \caption{}
  \label{fig:min-shap-deep}
  \end{subfigure}
  \begin{subfigure}[]{0.40\textwidth}
  \begin{center}
    \includegraphics[trim=45 00 45 00, clip, width=\textwidth]{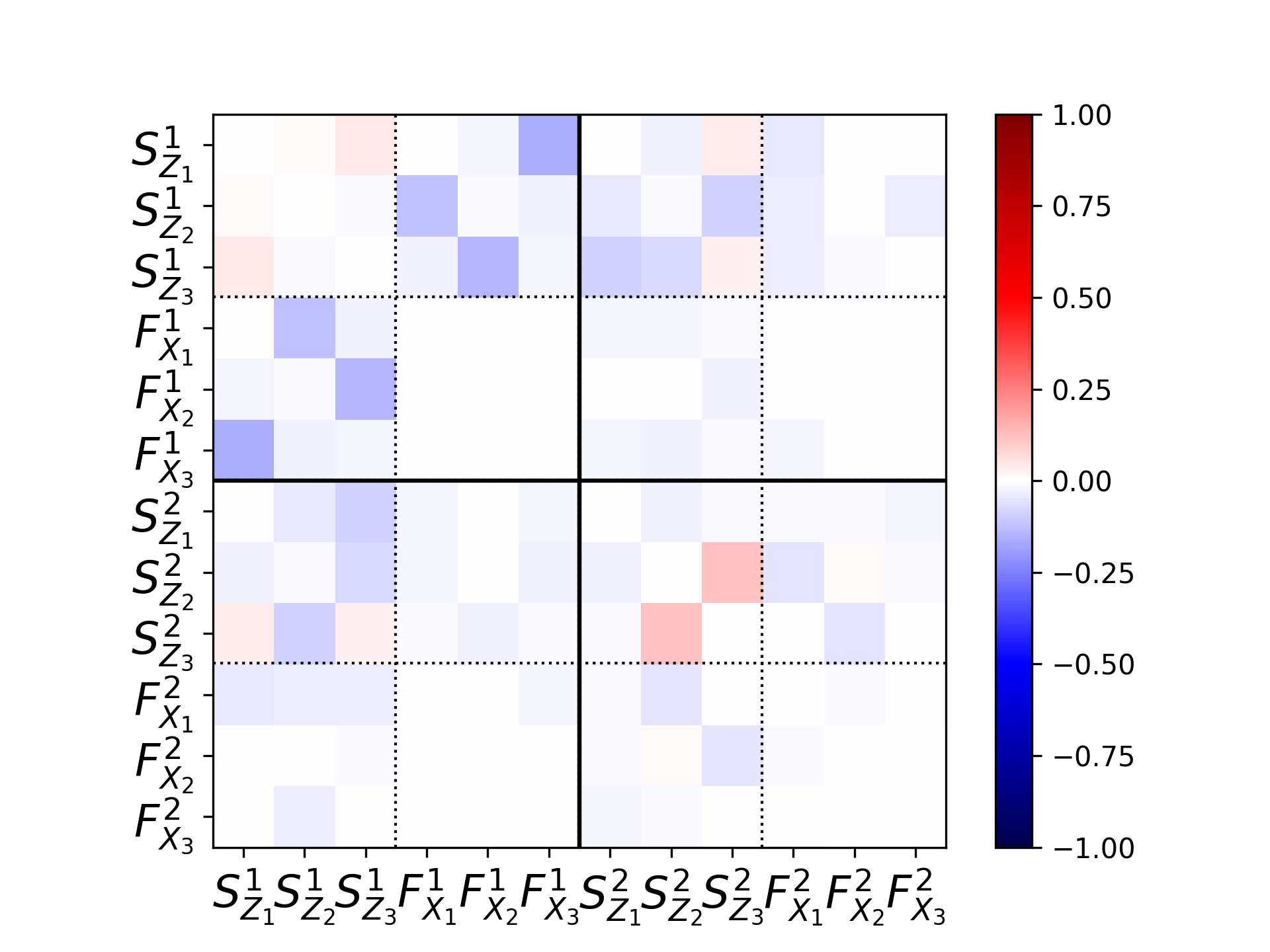}
  \end{center}
  \caption{}
  \label{fig:min-shap-delta}
  \end{subfigure}
\end{center}
  \caption{Correlation of Shapley values for the complete input of the trained DNN decoder for: (a) using the `Exact' explainer SHAP library~\cite{SHAPGit} and (b) and using the `DeepSHAP' approximation of the Shaley value. The `Exact' explainer works by considering all possible subsets of input features as explained in the main text and the `DeepSHAP' approximation corresponds to the method presented in ~\cref{subsec:num_shap_approx}. (c) Difference between of the correlation matrices shown in (a) and (b). Blue pixels represent where DeepSHAP causes an overestimation in the correlations, red pixels show an underestimation of the DeepSHAP method. The labels of each correlation matrix specify the input feature: $S$ for syndrome-, $F$ for flag-bit; the superscript labels the readout round and the subscript indicates the stabilizer measurement to which the respective bit corresponds.}
\label{fig:min-shap-corr}
\end{figure}

To test the validity of the DeepSHAP approximation implementation for the neural network decoders discussed before, we compare the Shapley values generated by DeepSHAP to an exact calculation of these Shapley values. For the exact calculation, we refer to Eq. (1) of the main text. The computation is based on iterating over all possible subsets of input values that are allowed to be used by the NN. The excluded bits are set to the average bit value. This implies that no retraining for each subset is performed, the whole analysis is done with a single trained NN.
The validity test is conducted for a non-recurrent, dense neural network decoder, that is trained to predict the logical bit flip correction for a Steane code state after two rounds of syndrome extraction. We choose this example, because it has the minimal number of input bits for which a decoder can predict a FT and nontrivial correction. The input data, that is given to the NN consists of 3 $Z$ stabilizer bits ($S_{Z_i}$) and 3 $X$ stabilizer flag bits ($F_{X_j}$) per cycle. This amounts to 12 input bits and correspondingly 12 Shapley values that are to be calculated.
The Shapley values are calculated using both the exact and the approximate method for $14.000$ validation data points. A set of $1.000$ random training samples is used for both approaches to estimate the reference values that are needed.
The~\cref{fig:min-shap-corr} (a,b) shows the correlations of Shapley values calculated using either the exact calculation (a) or the DeepSHAP approximation method (b). These correlation matrices show qualitatively similar patterns. To judge the quantitative difference the difference of the Shapley value correlations is also shown in~\cref{fig:min-shap-corr}. It seems as though the DeepSHAP method distributes more credit to hook error configurations. Overall, the qualitative pictures of the respective interpretations based on these correlation matrices align. Although differences of the DeepSHAP approximation from the exact calculation are to be expected, this test confirms that such differences do not change the deduced interpretation. In conclusion, we confirm that the DeepSHAP approximation can be used faithfully for the problem of neural-network decoding in our setting.

\section{Layer-Wise Relevance Propagation}%
\label{sec:LRP}
Layer-Wise Relevance Propagation~\cite{Bach2015, Montavon2019} (LRP) can be considered as a larger class of XAI methods, to which also the Shapley value approximation belongs. Generally, LRP methods assign credit to inputs of a NN. These scores can differ from the Shapley value approximation and depend on the way of how a relevance score is propagated through the network. Thereby, different heuristic rules can be used. In particular, the propagation that is described in~\cref{subsec:num_shap_approx} yields a Shapley value approximation. 
In the following parts, we explain the general LRP methodology and provide results for different LRP rules to investigate the interpretability of non-recurrent feed-forward (dense) NN decoders. As we will see,  these will provide a less resolved importance attribution and therefore less clear interpretations compared to the Shapley value method of the main text. By providing the latter comparison, we further motivate the utilization of the Shapley value approximation method over possible alternatives.

\subsection{Introduction to Layer-Wise Relevance Propagation}
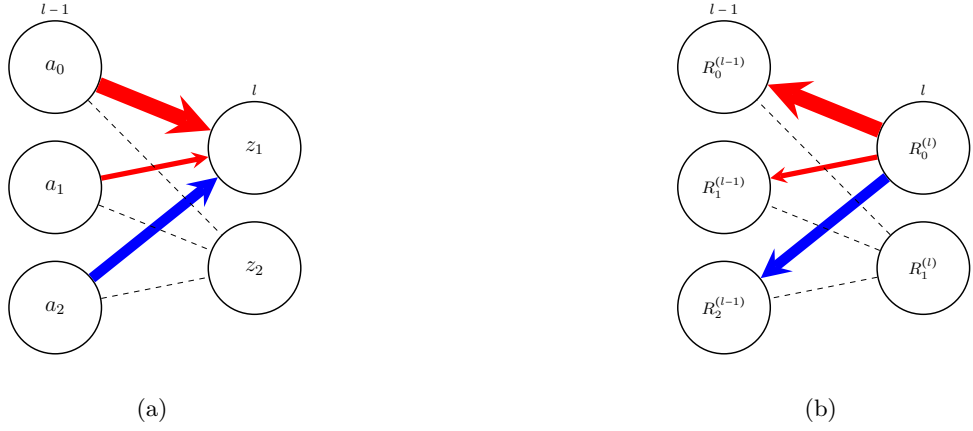
\begin{figure}[H]
\begin{center}
  \begin{subfigure}[]{0.49\textwidth}
  \begin{center}
    \scalebox{0.7}{
  \begin{tikzpicture}[node distance=0.5cm and 2cm]
    \tikzstyle{neuron} = [circle, draw=black, align=left, thick, minimum size=50pt];
    \tikzstyle{arrow} = [arrows = {-Latex[width'=0pt .5, length=10pt]}, thick];
    \tikzstyle{line} = [thick];
  
    \node[neuron, label={[above] \footnotesize $l-1$}] (j0) at (0, 0) {\large $a_0$};
    \node[neuron] (j1) [below=of j0] {\large $a_1$};
    \node[neuron] (j2) [below=of j1] {\large $a_2$};
    \node[neuron, label={[above] \footnotesize $l$}] (k0) [right=of j1, yshift=0.75cm] {\large $z_1$};
    \node[neuron] (k1) [below=of k0] {\large $z_2$};

    \draw [-stealth, line width=3mm, color=red] (j0) -- (k0);
    \draw [-stealth, line width=1mm, color=red] (j1) -- (k0);
    \draw [-stealth, line width=2mm, color=blue] (j2) -- (k0);
    \foreach \k in {1}
      \foreach \j in {0, 1, 2}
        \draw [dashed] (j\j) -- (k\k);
  \end{tikzpicture}}
  \end{center}
  \caption{}
  \label{fig:}
  \end{subfigure}
  \begin{subfigure}[]{0.49\textwidth}
  \begin{center}
    \scalebox{0.7}{
  \begin{tikzpicture}[node distance=0.5cm and 2cm]
    \tikzstyle{neuron} = [circle, draw=black, align=left, thick, minimum size=50pt];
    \tikzstyle{arrow} = [arrows = {-Latex[width'=0pt .5, length=10pt]}, thick];
    \tikzstyle{line} = [thick];
  
    \node[neuron, label={[above] \footnotesize $l-1$}] (j0) at (0, 0) {\small $\relv{l-1}{0}$};
    \node[neuron] (j1) [below=of j0] {\small $\relv{l-1}{1}$};
    \node[neuron] (j2) [below=of j1] {\small $\relv{l-1}{2}$};
    \node[neuron, label={[above] \footnotesize $l$}] (k0) [right=of j1, yshift=0.75cm] {$\relv{l}{0}$};
    \node[neuron] (k1) [below=of k0] {$\relv{l}{1}$};

    \draw [-stealth, line width=3mm, color=red] (k0) -- (j0);
    \draw [-stealth, line width=1mm, color=red] (k0) -- (j1);
    \draw [-stealth, line width=2mm, color=blue] (k0) -- (j2);
    \foreach \k in {1}
      \foreach \j in {0, 1, 2}
        \draw [dashed] (k\k) -- (j\j);
  \end{tikzpicture}}
  \end{center}
    \caption{}
  \label{fig:}
  \end{subfigure}
\end{center}
\caption{(a) Forward push, causing the activations $z_k=g\left(\sum_j a_j w_{jk}\right)$.  Here, the red arrows indicate positive contributions, whereas the blue arrows indicate negative contributions. The thickness of the arrows is proportional to the absolute contribution of that connection. (b) Relevance backward push: the relevance $R_0^{(l)}$ is redistributed to $R_j^{(0)}$ according to the relative contributions $a_j$ made to producing $z_1$. The relevance $R_j^{(0)}$ is summed over all backwards redistributions towards neuron $j$ of layer $l$.}
\label{fig:LRP_illustration}
\end{figure}

LRP works by backpropagating a conserved `relevance' quantity layer by layer through a network from the output to the input domain. This ansatz is generally motivated by the way a NN performs computations through updating the activation values of a layer based on the activation values of the preceding layer (forward pass), see~\cref{fig:LRP_illustration} (a). The systematic of relevance-backpropagation is illustrated for a pair of layers in~\cref{fig:LRP_illustration} (b). Backpropagation of the relevance score is generally done for individual input samples and depends on the specific NN state of all neurons as calculated in the forward pass. LRP operates on the principle that a neuron is considered relevant to producing a specific output in a neuron in the subsequent layer if it significantly contributed to the activation of that neuron.
The simplest relevance backpropagation rule is denoted LRP-$0$ rule and formulated as~\cite{Montavon2019}
\begin{equation}
\label{eq:LRP0}
  R^{(l)}_j = \sum_k \frac{w_{jk}a_{j}}{\sum_j w_{jk}a_{j}} R^{(l+1)}_k.
\end{equation}

Here $R^{(l)}_j$, $R^{(l+1)}_k$ are the respective relevance values of the $j$-th neuron in the $l$-th layer and the $k$-th neuron in the $(l+1)$-th layer. 
During the forward pass, the $j$-th neuron in layer $l$ contributes the input $w_{jk}a_j$ to the activation of the $k$-th neuron in layer $l+1$. The total activation of layer $l+1$ is $z_k = \sum_j w_{jk}a_k$, making the $j$-th neuron's relative contribution $\frac{w_{jk} a_k}{z_k}$. Relevance is then redistributed according to these relative activations.
It is assumed that $R_k^{(l+1)}$ is known. The relevance of the final layer is by convention set to the model output, $R_{\text{final}}=f(\vec x)$ or simply to $R_{\text{final}}=1$ when backpropagating a normalized relevance score. Note that non-linear activation functions are explicitly ignored. In this sense LRP linearly approximates the model.

The rule described in~\cref{eq:LRP0} is commonly referred to as the LRP-0 rule~\cite{Montavon2019} and is the simplest layer-wise relevance propagation rule. A common amendment to this rule to prevent numerical instability when $z_k=\sum_j w_{jk} a_j\to 0$ is to add a small numerical constant $\tilde\epsilon=\mathrm{sign}(z_k)\epsilon$ giving the LRP-$\epsilon$ rule~\cite{Montavon2019}:
\begin{equation}
  R^{(l)}_j = \sum_k \frac{w_{jk}a_{j}}{\tilde\epsilon + \sum_j w_{jk}a_{j}} R^{(l+1)}_k
\end{equation}

In certain situations, it may be desirable to treat positive activations differently from negative activations. To this end, several heuristic alternatives have been devised. A generalized backpropagation rule is given by~\cite{Montavon2019}
\begin{equation}
  R^{(l)}_j = \sum_k \frac{\rho(w_{jk})a_{j}}{\tilde\epsilon + \sum_j \rho(w_{jk})a_{j}} R^{(l+1)}_k,
\end{equation}
where $\rho$ is some transformation of the weight matrix. 

\begin{table}[H]
\begin{center}
  \begin{tabular}{ll}
  \toprule 
  Input domain & Rule\\
  \midrule
  \midrule
  Pixel intensities\newline $x_i\in[l, h]$, $l\le 0\le h$ 
    & $R_i = \sum_j \frac{x_i w_{ij} - l w^+_{ij} - h w^-_{ij}}{\sum_i x_i w_{ij} - l w^+_{ij} - h w^-_{ij}} R_j $\\
    Real values\newline $x_i\in\mathbb{R} $
    & $R_i = \sum_j \frac{w^2_{ij}}{\sum_i w^2_{ij}} R_j $\\
  \toprule 
  $w^+_{ij} = \min(0, w_{ij})$, $w^-_{ij}=\max(0, w_{ij})$
\end{tabular}
\end{center}
\caption{Different LRP rules, where the LRP-0 is augmented by a mapping of the weight matrix values $\rho$ that discriminates between positive and negative weights and treats them differently. In Ref.~\cite{Montavon2018}, these two example rules are introduced and motivated for problems with different input domains. }
\label{tab:input-domain-lrp-rules}
\end{table}
\begin{table}[H]
\begin{center}
  \begin{tabular}{ll}
    \toprule 
    Name & Transformation\\
    \midrule
    \midrule
    LRP-$\gamma$~\cite{Montavon2019} & $R^{(l)}_j = \sum_k \frac{(w_{jk} + \gamma w^+_{jk})a_{j}}{\tilde\epsilon + \sum_j (w_{jk} + \gamma w^+_{jk})a_{j}} R^{(l+1)}_k$\\
    LRP-$\alpha\beta$~\cite{Bach2015} 
    & $R^{(l)}_j = 
    \sum_k \left(
    \alpha\frac{w^+_{jk}a_{j}}{\tilde\epsilon + \sum_j w^+_{jk}a_{j}}
    -\beta\frac{w^-_{jk}a_{j}}{\tilde\epsilon + \sum_j w^-_{jk}a_{j}}
    \right) R^{(l+1)}_k$\\
    \toprule 
    $w^+_{ij} = \min(0, w_{ij})$, $w^-_{ij}=\max(0, w_{ij})$
  \end{tabular}
\end{center}
\caption{Summary of the LRP-$\gamma$ and LRP-$\alpha\beta$ rules as modifications of the LRP-$\epsilon$ rule. The parameters $\alpha,\beta,\gamma$ and $\epsilon$ are free ``interpretability-hyperparameters'' to be set or optimized upon.}
\label{tab:lrp_rules}
\end{table}

To backpropagate relevance into the input layer without biasing the explanation by direct multiplication with the input activations, special rules can be used. Note that using the LRP-$\epsilon$ rule for instance, zero input values could not be assigned relevance for instance, which is an obvious sign of bias. In~\cite{Montavon2018}, the rules shown in~\cref{tab:input-domain-lrp-rules} for backpropagation into the input layer are proposed.
Another recommendation of the same study is to use the LRP-$\alpha\beta$ rule with $\alpha=1,\beta=0$ (written LRP-$\alpha_1\beta_0$) when backpropagating in layers with ReLU activations, see~\cref{tab:lrp_rules} for the definition of this rule.
These heuristic backpropagation rules have been successfully applied in image recognition tasks~\cite{Bach2015, Montavon2018, Montavon2019}. Furthermore, a composition of rules within the process of  of backpropagation can be employed to explain, for instance, convolutional neural network based image classifiers~\cite{Montavon2019}. The goal of finding a good choice or composition of LRP rules is to obtain a expressive and unbiased interpretation capability.

\subsection{Using LRP to explain the functionality of a DNN decoder}%
\begin{figure}[H]
\begin{center}
  \begin{subfigure}[]{0.49\textwidth}
  \begin{center}
    \includegraphics[width=\textwidth]{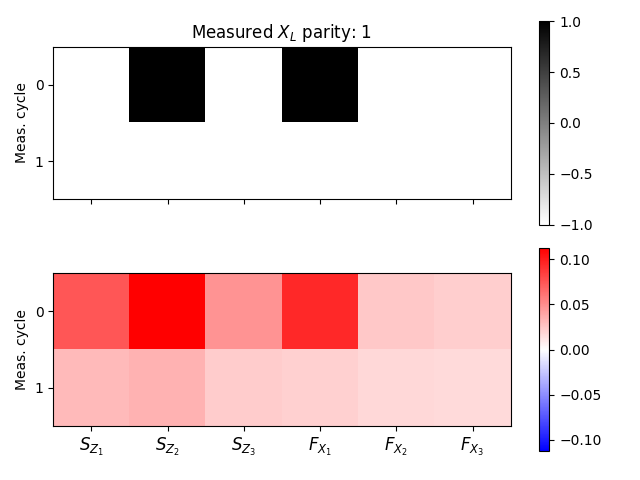}
  \end{center}
  \caption{}\label{fig:lrp-example-result}
  \end{subfigure}
  \begin{subfigure}[]{0.49\textwidth}
  \begin{center}
    \includegraphics[width=\textwidth]{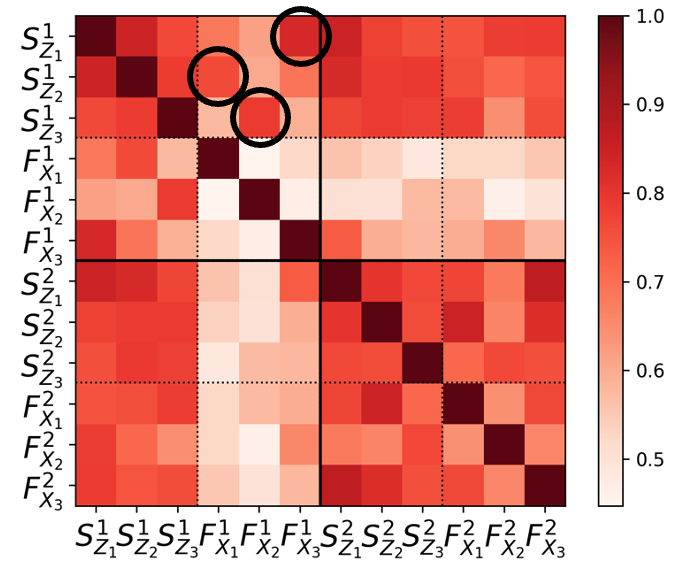}
  \end{center}
  \caption{}\label{fig:lrp-corr}
  \end{subfigure}
\end{center}
  \caption{(a) Exemplary saliency map using the LRP-$\alpha_1 \beta_0$ and the pixel intensities input layer rule. The top image shows the input data where a hook error was measured. The bottom shows the explanation produced using LRP. The relevant inputs are the most important features of the explanation, but there is still a lot of noise. (b) Correlation matrix of LRP on the validation dataset for the minimal data 2 cycle DNN decoder. The relatively larger correlations between stabilizer bits and flag bits (encircled squares) in the same measurement round are a sign that the network learned to identify hook error signatures.}
\label{fig:lrp_result}
\end{figure}

\label{sub:Using LRP to explain a DNN Steane Decoder}
Although LRP can be extended to LSTM units~\cite{Montavon2018}, we do not explore this as part of this work. For this reason, LRP is only tested using a DNN decoder that was trained to fault-tolerantly decode up to two syndrome measurement cycles.
To this end, a data set of $100.000$ samples of two stabilizer measurement cycles of the Steane code has been used to train a DNN network. The network has been trained to detect logical bit flip errors, and it only receives information about $X$-stabilizer plaquette flags and $Z$-stabilizer syndrome data. The training set has been split into a training set of $90.000$ samples and $10.000$ validation samples. The network has then been trained for 330 epochs and passed the deterministic fault placing benchmark after 173 epochs. The latter proves the FT decoding capability of the NN explicitly. LRP relevance scores are then calculated for all $10.000$ training samples. 
We present an analysis for the LRP-$\alpha_1\beta_0$ rule, defined in~\cref{tab:lrp_rules} combined with the pixel intensities rule of~\cref{tab:input-domain-lrp-rules} for the final backpropagation step to the input layer. We have chosen this method as a well-working LRP variant out of many that we have tried out. In~\cref{fig:lrp_result} (a) an example of a relevance computation is shown. This example is a specific hook error configuration with a non-trivial measurement of $F_{X_1}$ and  $S_{Z_2}$. It is the same example as illustrated in Fig. 5 (b,c) of the main text where it is also explained in  detail. Note that the relevance score attributions can be compared to Fig. 4 of the main text, where the same analysis is done based on the Shapley value approximation as a relevance score.

The LRP relevance map most strongly highlights the measured ancilla and flag flips, but we find that it also assigns relevance to the neighboring ancilla bits. The fact that these other ancilla bits are not flipped should not be particularly relevant to indicating that a hook error occurred, indicating that the resolution of the LRP method is not very high. There is a quite high background of relevance score on all input bits. Both of these shortcomings, visible in~\cref{fig:lrp_result} (a) can be compared with the counterpart of using the Shapley approximation method, shown in Fig. 4 of the main text. In general, we find that relevance mappings using this LRP rule are noisy and not particularly satisfying in their explanatory quality.

To obtain a global picture of what LRP deems relevant for this network, the correlation between all 12 input features for the $10.000$ LRP samples is calculated. This correlation matrix, analogous to the correlation matrices introduced in the main text for the Shapley score, is shown in~\cref{fig:lrp_result} (b). Here, some basic features of what the network has learned can be distinguished. When looking specifically at the correlations between the $S^1_{Z_i}$ and $F^1_{Z_j}$ for $j=i+1\mod 3$, one can make out that the hook error signatures, described in the main text, are seen as relevant indicators of logical errors. It can also be observed that the $F^1_{X_i}$ flags tend not to correlate strongly with any other features. Other than this, it is hard to see qualitative features of what the network has learned in this correlation matrix.
We explored the use of some other LRP rules like the LRP-$0$, the LRP-$\epsilon$, and the LRP-$\alpha_2\beta_1$ rule for the trained and FT feed-forward NN decoder. These LRP did not result in particularly better and clearer interpretability results and are not presented here extensively. 


\section{Sequential Look-Up-Table Decoder}%
\label{sec:Sequential-Look-Up-Table-Decoder}
To compare the performance of the RNN decoders used in the main text to a reference decoder, the sequential look-up-table (SeqLUT) decoder is introduced as well. It is a FT Pauli-frame tracking look-up-table (LUT) based decoder, able to decode the syndrome flag volume of a sequence of repeated stabilizer readouts by  applying a fixed decoding rule (the LUT) to a moving window of two consecutive rounds of syndrome and flag measurements. In the following part, we explain the decoding logic of the SeqLUT in more detail, including how the flag information is used to correct for hook errors.

\begin{figure}[H]
\begin{center}
  \begin{tikzpicture}[]
    \tikzstyle{layer} = [draw=black, align=left, fill=white]
    \node[layer] (s1) at (0, 0) {$\delta \vec s(0)$\\ $\vec f(0)$};
    \node[layer] (s2) [right=of s1]{$\delta \vec s(1)$\\ $\vec f(1)$};
    \node[layer] (s3) [right=of s2] {$\delta \vec s(2)$\\ $\vec f(2)$};
    \node[layer] (s4) [right=of s3] {$\delta \vec s(3)$\\ $\vec f(3)$};
    \node[] (s5) [right=of s4] {};


    \node[] (head) [below=of s2]{Head};
    \draw[->, thick] (head) -- (s2);
    \node[] (tail) [below=of s1]{Reference};
    \draw[->, thick] (tail) -- (s1);

    \draw[->] (s1) -- (s2);
    \draw[->] (s2) -- (s3);
    \draw[->] (s3) -- (s4);
    \draw[dashed] (s4) -- (s5);
    
    %
  \end{tikzpicture}
\end{center}
\caption{Schematic representation of the reference and head pointer of the sequential look-up-table decoder that picks the data used for sequential decoding. The entire decoding is done by sliding both pointers over the time series of syndrome-flag data. }%
\label{fig:seqlut-pointers}
\end{figure}
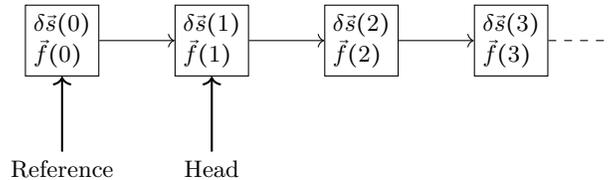
\begin{figure}[H]
\begin{center}
  \begin{subfigure}[]{0.45\textwidth}
  \begin{center}
  \resizebox{0.9\textwidth}{!}{%
  \begin{quantikz}
    \qw{} & \qw{1} & \targ{} &\qw{} &\qw{} &\qw{}&\qw{} & \qw{} & \qw{}&\qw{}&\qw{}
    \\
    \qw{} & \qw{2} &\qw{} &\qw{}& \targ{} &\qw{}&\qw{}&\qw{} & \qw{} & \qw{}&\qw{}
    \\
    \qw{} & \qw{3} &\qw{}&\qw{} &\qw{} & \targ{}&\qw{}&\qw{} & \qw{} & \qw{}&\qw{}
    \\
    \qw{} & \qw{4} &\qw{} &\qw{}&\qw{} &\qw{} &\qw{} & \targ{}&\qw{} & \qw{} & \qw{}
    \\
    \qw{\ket{+}} &\qw{}& \ctrl{-4}& \ctrl{1}& \ctrl{-3}\qw{\encirc{1}}& \ctrl{-2}\qw{\encirc{2}} &\ctrl{1}\qw{\encirc{3}}&\ctrl{-1} & \qw{} & \meterD{X} & \cw{}
    \\
    \qw{\text{Flag}\ \ket{0}} &\qw{}& \qw{} & \targ{} & \qw{}& \qw{}& \targ{} & \qw{} & \qw{} & \meterD{Z} & \cw{}
  \end{quantikz}%
  }
  \end{center}
  \caption{}
  \end{subfigure}
  \begin{subfigure}[]{0.45\textwidth}
  \begin{center}
  \resizebox{0.9\textwidth}{!}{%
  \begin{quantikz}
    \qw{} & \qw{1} & \control{} &\qw{} &\qw{} &\qw{}&\qw{} & \qw{} & \qw{} &\qw{} &\qw{}
    \\
    \qw{} & \qw{2} &\qw{} &\qw{}& \control{}&\qw{}&\qw{}&\qw{} & \qw{} & \qw{} &\qw{}
    \\
    \qw{} & \qw{3} &\qw{} &\qw{}&\qw{} & \control{}&\qw{}&\qw{} & \qw{} & \qw{}&\qw{}
    \\
    \qw{} & \qw{4} &\qw{} &\qw{}&\qw{}&\qw{} &\qw{} & \control{}&\qw{} & \qw{} & \qw{}
    \\
    \qw{\ket{+}} &\qw{}& \ctrl{-4} & \ctrl{1} & \ctrl{-3}\qw{\encirc{1}}& \ctrl{-2}\qw{\encirc{2}}&\ctrl{1}\qw{\encirc{3}} &\ctrl{-1} & \qw{} & \meterD{X} & \cw{}
    \\
    \qw{\text{Flag}\ \ket{0}} &\qw{}& \qw{} & \targ{} & \qw{}& \qw{}& \targ{} & \qw{} & \qw{} & \meterD{Z} & \cw{}
  \end{quantikz}%
  }
  \end{center}
  \caption{}
  \label{fig:f1ftec-z-plaq-meas}
  \end{subfigure}
\end{center}
  \caption{Circuitry for the flagged measurement of an $X$ [$Z$] type, weight-4 plaquette measurement of the Steane code in (a) [(b)]. In particular, the positions of possible CNOT errors are indicated by numbers that cause a non-trivial flag measurement.}
\label{fig:f1ftec-meas}
\end{figure}
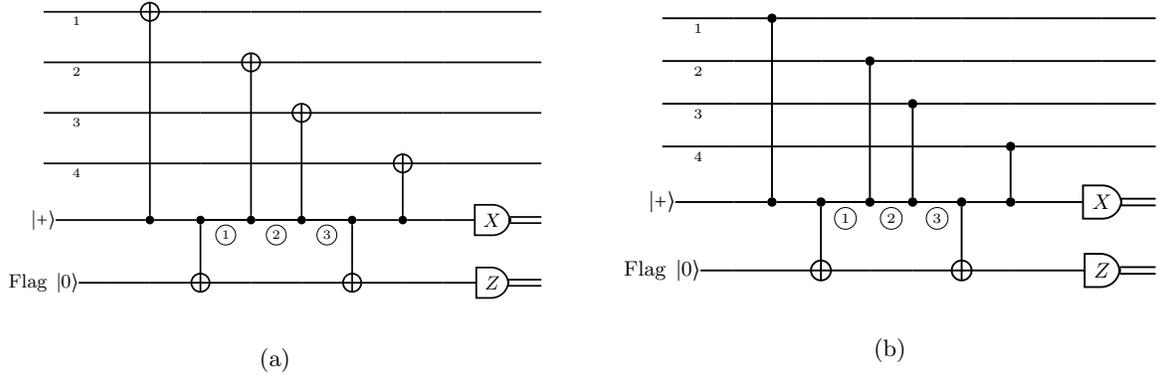
\begin{table}[h]
\begin{tabular}{ccccc}
  \toprule
  \toprule
  Flagged circuit & Ancilla error & Data Error & Stabilizer syndrome & Correction \\
  \midrule
  $S_{1234}$ & $E_1$ & $P_2P_3P_4=P_1$ & 100 & $P_1$ \\
  & $E_2$ & $P_3P_4$ & 010 & $P_5 \to P_3P_4$\\
  & $E_3$ & $P_4$ & 101 & $P_4$ \\
  \midrule
  $S_{2356}$ & $E_1$ & $P_3P_4P_5=P_2$ & 110 & $P_2$ \\
  & $E_2$ & $P_5P_6$ & 001 & $P_7 \to P_5P_6$\\
  & $E_3$ & $P_6$ & 011 & $P_6$ \\
  \midrule
  $S_{3467}$ & $E_1$ & $P_4P_6P_7=P_3$ & 111 & $P_3$ \\
  & $E_2$ & $P_6P_7$ & 010 & $P_5 \to P_6P_7$\\
  & $E_3$ & $P_7$ & 001 & $P_7$ \\
  \toprule
  \toprule
\end{tabular}
\caption{Correction table for the correction mode FLAG, i.e. the alteration of the standard weight-one correction. The numbering of qubits and stabilizers corresponds to Fig. 2 (a) of the main text. 'Flagged circuit' refers to the stabilizer for which a non-trivial flag bit was measured.
For every flagged circuit the three possible errors $E_i$ on the ancilla bit are considered at the encircled location $i=1,2,3$ in~\cref{fig:f1ftec-meas}.
These will cause different data errors due to the error propagation. Assuming a single plaquette measurement circuit flags, the regular stabilizer look-up-table will be modified to correct a weight-two data qubit error if an error has occurred at location $i=2$. How the correction changes depends on the measurement order of the plaquette. Here, an execution order of entangling gates of the flagged readout circuit of the stabilizer $S_{ijkl}$ is assumed to be $i\to j \to k \to l$.}
\label{tab:LUT}
\end{table}

The SeqLUT decoder works by sliding with a `head' pointer and a `reference' pointer over a sequence of syndrome measurement cycles, schematically shown in~\cref{fig:seqlut-pointers}. At the initialization of the code, the reference pointer is set to $t=0$ and the head pointer to $t=1$, and a `signal' variable is set to NONE. This signal variable is used to activate different types of correction modes. The head can detect three types of signals depending on the syndrome and flag information of the measurement cycle the head is currently at:
\begin{enumerate}
  \item \textbf{NONE} if no syndrome increments occur and no flags occur
  \item \textbf{ERROR} if no flags occur but a syndrome increment is detected
  \item \textbf{FLAG} if a flag is detected.
\end{enumerate}
The head pointer is looped over the data with the signal at NONE until one of the following events occur.
\begin{enumerate}
  \item If the signal is NONE and a flag is measured at the head, the signal is set to FLAG and the head is incremented by one.
  \item else, if the signal is NONE and the head and reference have a mutual increment, the signal is set to ERROR, and the head is incremented by one.
  \item otherwise, the signal is set to NONE and the head is incremented by one.
\end{enumerate}

If the signal is at FLAG or ERROR, an appropriate error is determined using the LUT scheme. If no flag is triggered, this correction corresponds to the standard weight-one correction that uniquely revokes the violation of the stabilizers. If a flag was triggered, a hook error might have occurred if there was a circuit level propagating error at location $2$ in the readout circuits in~\cref{fig:f1ftec-meas}. Depending on the complementary syndrome values this case can be detected, and the standard correction is changed to its logical complement, which is given by a weight-two correction. The precise LUT flag rules are given in~\cref{tab:LUT}.
The reference is then set to the current head, and the head is incremented by one. This procedure is repeated until the end of the sequence.

The SeqLUT decoder can be used to analyze the same dataset used for the analysis of the LSTM based decoders. The SeqLUT decoder passes the deterministic fault placing benchmark for both decoding bases. This confirms its fault tolerance explicitly.
The infidelity curves provided by employing the SeqLUT decoder can be fitted with the infidelity $\mathcal I(t)$ defined in~\cref{eq:infidelity}. This is shown in Fig. 2 (d) in the main text for logical bit-flip decoding. A comparison of the performance of both RNN decoders and the SeqLUT decoder in both measurement bases is given in the same figure.

\section{Benchmarking of the decoders}
In the main text, the presented decoders are benchmarked by different means based on simulated data. Details regarding this benchmarking are given in the following sub-sections.
\subsection{Fitting the logical error rate per round}
Assuming each measurement cycle has a probability $p_L$ to introduce a logical flip, the logical fidelity $\mathcal F(t)$ is a measure of how many logical qubits from an ensemble have had an even number of logical flips after $t$ measurement cycles~\cite{Baireuther2019}, 
\begin{equation}
  \mathcal F(t) = \frac{1}{2} + \frac{1}{2} (1-2p_L)^{t-t_0}.
\end{equation}
A derivation for this is given by O'Brien~\textit{et al.}~\cite{OBrien2017}. The logical infidelity $\mathcal I(t)$ is given by
\begin{equation}\label{eq:infidelity}
  \mathcal{I}(t) = 1 - \mathcal{F}(t) = \frac{1}{2} - \frac{1}{2} (1-2p_L)^{t-t_0}.
\end{equation}
To determine $p_L$ and $t_0$, a least-square fit is performed. The parameter $t_0$ can capture logical state preparation and measurement (SPAM) errors.
In order to benchmark the FT of a decoder, $p_L$ is inferred from this exponential fit for different values of the physical error rate $p_{ph}$ of the circuit level noise model. The FT of a decoder would be indicated, for the distance-3 QEC code under consideration, by scaling of the logical error rate as
\begin{equation}
    p_L\propto p_{ph}^2+\mathcal{O}(p_{ph}^3),
\end{equation}
for low enough values of $p_{ph}$. Correspondingly, FT can be confirmed heuristically by performing a fit of the obtained logical error rates to $a\cdot p_{ph}^b$ restricting the interval of physical error rates to small enough values.

\subsection{Uncertainty estimation and the Wilson interval}
\label{subsec:wilson}
The Wilson interval~\cite{WilsonInterval, WilsonOriginal, BerryPhaseWilsonInterval2014} is a method for estimating a confidence interval (CI) to characterize statistical uncertainty of samples drawn from a binomial distribution 
\begin{equation}
  \mathcal{B}(n, k, p)=\begin{pmatrix} n \\ k\end{pmatrix}p^k(1-p)^k. 
\end{equation}
It is able to capture the statistical uncertainty more accurately then the standard variance estimator in the case of low $p$ and very few samples $k$ in one class.
The goal is to find an estimate with statistical uncertainty for $p$ from an empirical sample $\{x_i\}_i$ for $x_i\in\{0, 1\}$ assumed to be drawn from $\mathcal B(n, k, p)$. We would like to define an interval of possible values such that the true value of $p$ lies within the CI with probability of $1-\epsilon$ for some arbitrary $\epsilon > 0$. The Wilson interval defines a CI around the mean of the empirical sample
\begin{equation}
  \hat p = \langle x_i\rangle,
\end{equation}
such that $p\in[p_{\mathrm{min}}, p_{\mathrm{max}}]$ with probability $1-\epsilon$ for
\begin{equation}
  p_{\mathrm{max}, \mathrm{min}}(\hat p; \epsilon, n) = \left(
  \hat p + \frac{z^2_{\epsilon/2}}{2 N} \pm z_{\epsilon/2}\sqrt{
    \frac{\hat p(1-\hat p)}{N}  + \frac{z^2_{\epsilon/2}}{4N^2} 
  }
  \right)  \left( 1 + \frac{z^2_{\epsilon/2}}{N}
  \right)^{-1},
\end{equation}
where $\epsilon$ defines our CI, $N$ is the number of observations in the sample $\{x_i\}_i$, and $z_{\epsilon/2}$ is the quantile function of the normal distribution~\cite{cowan1998statistical}.

Following the example of Dauphin \textit{et al.}~\cite{BerryPhaseWilsonInterval2014}, $p_\mathrm{min}$ and $p_\mathrm{max}$ are cut off to diminish fringe effects for
\begin{equation}
  \begin{split}
    p_{\mathrm{min}} = 0 &\text{ if } \hat p = l/n,\ l \in \{0, 1, 2\}, \\
    p_{\mathrm{max}} = 1 &\text{ if } \hat p = l/n,\ l \in \{n-2, n-1, n\}, \\
  \end{split}
\end{equation}
including $l=3,$ and $l=n-3$ respectively if $n>40$. For convenience, to obtain symmetric error bars the CI is symmetrized around $\hat p$ by choosing the maximal deviation
\begin{equation}
  \sigma_{\hat p}=2\cdot\max\{\abs{\hat p - p_\mathrm{max}}, \abs{\hat p - p_\mathrm{min}}\},
\end{equation}
such that $p_\mathrm{true}\in p\pm \sigma_{\hat p}$ with probability $P \ge 1 - \epsilon$.
To obtain the error bars shown in the main text, the probability of the CI is fixed to the $1\sigma$ CI of the normal distribution, i.e. $z^2_{\epsilon/2}=1$.

\end{document}